\documentclass[lettersize,journal]{IEEEtran}
\usepackage{amsmath,amsfonts}
\usepackage{amssymb}
\usepackage{algorithmic}
\usepackage[caption=false,font=footnotesize,labelfont=sf,textfont=sf]{subfig}
\usepackage{stfloats}
\usepackage{url}
\usepackage{graphicx}
\usepackage{cite}
\usepackage{float}
\usepackage{subfig}
\usepackage{multirow}
\usepackage{color}
\usepackage{makecell}
\usepackage{booktabs}
\usepackage[ruled,linesnumbered]{algorithm2e}
\usepackage[colorlinks,citecolor=blue,linkcolor=blue]{hyperref}
\newtheorem{corollary}{Corollary}
\newtheorem{theorem}{Theorem}

\newtheorem{property}{Property}
\makeatletter

\newcommand{\Rmnum}[1]{\expandafter\@slowromancap\romannumeral #1@}
\makeatother

\def\BibTeX{{\rm B\kern-.05em{\sc i\kern-.025em b}\kern-.08em
    T\kern-.1667em\lower.7ex\hbox{E}\kern-.125emX}}
\usepackage{balance}
\begin{document}

\title{{Communications under Bursty Mixed Gaussian-impulsive Noise: Demodulation and Performance Analysis}}

\author{Tianfu Qi, \emph{Graduate Student Member, IEEE}, Jun Wang, \emph{Senior Member, IEEE}, Zuxue Zhao
\thanks{Tianfu Qi, Jun Wang are with the National Key Laboratory of Wireless Communications, University of Electronic Science and Technology of China, Chengdu 611731, China (e-mail: 202311220634@uestc.edu.cn)}
\thanks{Zexue Zhao is with the New Technology Research Department, China Ship Research and Development Academy, Beijing 100000, China (e-mail: zhaozexue@126.com).}
}

\markboth{Qi \MakeLowercase{\textit{et al.}}: Communications under Bursty Mixed Gaussian-impulsive Noise: Demodulation and Performance Analysis}%
{Shell \MakeLowercase{\textit{et al.}}: A Sample Article Using IEEEtran.cls for IEEE Journals}
\maketitle

\begin{abstract}
This is the second part of the two-part paper considering the communications under the bursty mixed noise composed of white Gaussian noise and colored non-Gaussian impulsive noise. In the first part, based on Gaussian distribution and student distribution, we proposed a multivariate bursty mixed noise model and designed model parameter estimation algorithms. However, the performance of a communication system will significantly deteriorate under the bursty mixed noise if a conventional signal detection algorithm with respect to Gaussian noise is applied. To address this issue, in the second part, we leverage the probability density function (PDF) to derive the maximum likelihood (ML) demodulation methods for both linear and nonlinear modulations, including M-array PSK (M-PSK) and MSK modulation schemes. We analyze the theoretical bit error rate (BER) performance of M-PSK and present close-form BER expressions. For the MSK demodulation based on the Viterbi algorithm, we derive a lower and upper bound of BER. Simulation results showcase that the proposed demodulation methods outperform baselines by more than 2.5dB when the BER performance reaches the order of magnitude of $10^{-3}$, and the theoretical analysis matches the simulated results well.
\end{abstract}

\begin{IEEEkeywords}
bursty mixed noise, maximum likelihood demodulation, theoretical performance, Viterbi algorithm
\end{IEEEkeywords}

\section{Introduction}

The non-Gaussian impulsive noise (IN), which has outliers with short duration and large amplitude, exists in many practical environments such as underwater acoustic communication \cite{paper2}, man-made devices, wireless digital video
broadcasting terrestrial (DVB-T), LTE in urban environments \cite{paper1}, etc. The IN can lead to significant performance deterioration of communication systems. Therefore, it is necessary to design the optimal detector based on the model well describing the noise statistics.

There is much literature considering the signal detection under IN recently \cite{paper2,paper20,paper22,paper23}. For instance, Middleton class A/B/C (MCA/B/C) models are proposed based on the physical characteristic and noise spectrum features \cite{paper7}. The MCA/B/C models are heavy-tail distributions and can accurately describe the non-Gaussian impulsive noise in various scenarios but with quite complicated probability density function (PDF). Shao et al. utilize the symmetric $\alpha$ stable (S$\alpha$S) distribution to approximately model the IN, but S$\alpha$S distribution does not have close-form PDF except for special cases including Cauchy distribution ($\alpha=1$) and Gaussian distribution ($\alpha=2$) \cite{paper8}. Gonzalez et al. propose a maximum likelihood (ML) detector termed as `myriad detector' by assuming the noise follows the Cauchy distribution \cite{paper18}. Furthermore, many improved versions are subsequently designed such as weighted myriad filter, recursive weighted myriad filter, etc \cite{paper24,paper25}.

All of the mentioned algorithms are with respect to the white IN for which noise samples are mutually independent and identically distributed. However, the IN usually has memory and then, several consecutive noise samples simultaneously have quite large amplitude \cite{paper4}. In this case, the energy of IN will mainly concentrate on a short duration, which leads to bursty symbol errors in a communication system. To solve this problem, Mahmood et al. adopted the $\alpha$ sub-Gaussian ($\alpha$SG) distribution to model the bursty IN \cite{paper9}. For their considered $\alpha$SG model, the random vector has identically $\alpha$ stable distributed components. As $\alpha$SG model has no analytical PDF, some sub-optimal detectors are derived in \cite{paper17}. By leveraging the Markov process and mixture Gaussian distribution to model the noise, Dario Fertonani et al. utilized the Bahl–Cocke–Jelinek–Raviv (BCJR) algorithm to implement a recursive detector to maximize a posteriori probability \cite{paper26}. However, as the mixture Gaussian model does not admit heavy tail, it is far from the real noise. The resulting method is also sub-optimal in practical applications.

However, the aforementioned noise model does not consider the white Gaussian noise (WGN) that is ubiquitous in a communication system. Sureka et al. propose an approximated white mixed noise model and derive the corresponding ML metric for BPSK demodulation \cite{paper14}. With the aid of this white mixed noise model, the Viterbi algorithm (VA) of MSK demodulation is considered in \cite{paper20}. To the best of authors' knowledge, there is no effective ML demodulation method reported with respect to the mixed noise of IN with memory and WGN due to the lack of concise and analytical noise model.

To address the aforementioned problem, in the first part of this two-part paper, we have proposed a multivariate model termed as `GS model' consisting of Gaussian distribution and student distribution to model the bursty mixed noise. The PDF admits a close-form expression, which makes it convenient to derive the optimal detector and perform theoretical analysis. In the second part of this two-part paper, we further leverage the GS model to design the ML demodulation algorithm for both linear modulation, such as M-array PSK (M-PSK), and nonlinear modulation, such as MSK scheme. Furthermore, the demodulation theoretical performance is systematically analyzed in various scenarios. Our contributions are summarized as follows,
\begin{enumerate}
\item{First, based on the proposed GS noise model and Markov property, we derive the ML metrics for both M-PSK and MSK schemes in the bursty mixed noise. Comparing with existing methods, simulation results show that the obtained ML demodulation algorithm can achieve better demodulation performance. The performance gain could reach at least 2.5dB when the bit error rate (BER) is of the order of magnitude of $10^{-3}$.}
\item{Second, the theoretical BER performance of M-PSK is investigated. We obtain the asymptotic performance expression with closed form for BPSK and QPSK schemes. Moreover, analytical lower and upper bounds for M-PSK (M>4) are also proposed. Numerical results demonstrate the theoretical BER converges to the simulated BER in the high signal-to-noise ratio (SNR) region and the bounds are also tight.}
\item{Last but not least, we present a lower bound and an upper bound of BER performance for VA-based MSK demodulation in the bursty mixed noise, respectively. We approximate the BER by mainly considering the error events with the largest probability. Simulations verify that our bounds are rather tight with the simulated BER, especially for the cases that the impulsiveness of the bursty mixed noise is prominent.}
\end{enumerate}

The remainder of the paper is organized as follows. In section \ref{statisticalmodel}, we briefly describe our proposed bursty mixed noise model and its key properties for the sake of completeness. In section \ref{application}, the ML demodulation algorithm and theoretical error probability under bursty mixed noise for BPSK modulated signals are investigated. Furthermore, the ML metric is designed for MSK signals and theoretical performance bounds are proposed in section  \ref{section_demodulation_MSK}. Numerical results are presented in section \ref{simulation}. Finally, the whole paper is concluded in section \ref{conclusion}.

$\mathbf{Notations}$: A random variable (RV) and its samples are separately presented by unbolded uppercase and lowercase letters, e.g., $X$ and $x$. Specifically, we denote the $i$-th sample vector as $\mathbf{x}_i=[x_i(1),x_i(2),\cdots,x_i(p)]^\top$. We use the notations `$\Sigma_{i,j}$' and `$\Sigma(i,j)$' to denote the block matrix and the element of the $i$-th row and $j$-th column of $\Sigma$, respectively. For instance, the matrix $\Sigma$ is partitioned as follows:

\begin{equation}
\Sigma=
\left[
\begin{array}{cc}
  \mathbf{A} & \mathbf{B} \\
  \mathbf{C} & \mathbf{D}
\end{array}
\right],
\end{equation}
Then, we denote $\mathbf{C}$ by $\Sigma_{2,1}$ and $\mathbf{B}$ by $\Sigma_{1,2}$. We utilize the notation `$P(\cdot)$' to represent the probability of a set. $\mathbb{E}_{\mu_{X}}(X)$ is the expectation of RV $X$ where $\mu_{X}$ is the probability measure of $X$. The notation `$\Vert\cdot\Vert$' denotes the Frobenius norm. $\mathbb{R}^{p}$, $\mathbb{Z}^{+}$ and $\mathbb{S}^{p}_{++}$ separately represent the $p$-dimensional domain of real numbers, positive integer set and the set of $p\times p$ positive definite matrix. $\det(\cdot)$ is the determinant operation. $\mathbf{e}_p$ denotes the $p$-dimensional all-one vector. Finally, we denote the $p$-order identical matrix by $\mathbf{I}_p$ and the Gamma function to be $\Gamma\left(a\right)=\int_{0}^{+\infty}t^{a-1}\exp(-t)dt$.

\section{Noise model}\label{statisticalmodel}
We consider the bursty mixed noise consisting of WGN and bursty IN and assume the WGN and IN are mutually independent since they stem from different sources. In this section, we briefly redescribe the noise model, which we have thoroughly presented in the first part of this two-part paper.

The PDF of GS model is as follows,
\begin{equation}\label{bursty_mixed_noise_model}
f_M(\mathbf{n})=\rho k_1\exp\left(-\frac{\Vert\mathbf{n}\Vert^2}{4\gamma_g^2}\right) +\frac{(1-\rho)k_2}{\left(1+\Vert\Sigma^{-1/2}\mathbf{n}\Vert^2/\alpha\right)^\frac{\alpha+p}{2}},
\end{equation}
where $k_2=\Gamma(\frac{\alpha+p}{2})/\left(\Gamma(\frac{\alpha}{2})(\alpha\pi)^{p/2}\sqrt{\det(\Sigma)}\right)$ and $k_1=(2\sqrt{\pi}\gamma_g)^{-p}$ and are normalization factors for WGN and bursty IN components, respectively. For convenience, we denote $\Sigma=2\gamma_s^2\tilde{\Sigma}$ and thus, the diagonal elements of regularized covariance matrix $\tilde{\Sigma}$ are 1.

The parameters in \eqref{bursty_mixed_noise_model} are $\alpha$, $\gamma_g$, $\gamma_s$, $\rho$, $\Sigma$ and $p$. Similar to S$\alpha$S distribution, $\alpha$ is the characteristic parameter to represent the heaviness of the PDF tail. For practical communication applications, we herein consider $\alpha\in(0,2)$, which corresponds to $\alpha$ stable distribution. $\gamma_g\in(0,+\infty)$ and $\gamma_s\in(0,+\infty)$ are the scale parameters of WGN and bursty IN, respectively. $\rho\in[0,1]$ is the weight parameter to adjust the mainlobe and tail section of the PDF, which describes the ratio of WGN and bursty IN. $\Sigma\in\mathbb{S}^{p}_{++}$ and $p\in\mathbb{Z}^{+}$ are the covariance matrix and memory order of IN, respectively. Consequently, we have $\Sigma^{-1}=(\Sigma^{-1/2})^\top\Sigma^{-1/2}$ by the Cholesky decomposition and $\Sigma^{-1/2}$ is an upper triangular matrix.

Obviously, the PDF expression \eqref{bursty_mixed_noise_model} is analytical. Hence, numerical integral in S$\alpha$S model and infinite sum in MCA/B/C model are avoided. Meanwhile, as presented in the first part of this two-part paper, the parameters of this noise model can be effectively estimated based on the PDF and characteristic function (CF). Then, we present two important properties which will be used in what follows. The detailed proofs of them can be found in the first part of the two-part paper.
\begin{property}\label{property_2}
The conditional PDF of $f_M(\mathbf{n})$ is
\begin{align}\label{conditional_PDF}
f_M(n|\mathbf{n}_1)=&\frac{1}{f_M(\mathbf{n}_1)}\Bigg[\rho k_1\exp\left(-\frac{n^2+\Vert\mathbf{n}_1\Vert^2}{4\gamma_g^2}\right) \nonumber\\
&+\frac{(1-\rho)k_2}{\left(1+\left(\Vert\Sigma_{1,1}^{-1/2}\mathbf{n}_1\Vert^2+(n-\mu)^2/\sigma\right)/\alpha\right)^\frac{\alpha+p}{2}}\Bigg],
\end{align}
where $\mathbf{n}=[\mathbf{n}_1^\top,n]^\top$, $\mu=\mathbf{n}_1^\top\Sigma_{1,1}^{-1}\Sigma_{1,2}$, $\sigma=\frac{\det(\Sigma)}{\det(\Sigma_{1,1})}$ and $\Sigma$ is partitioned to
\begin{equation}
\Sigma=
\left[
\begin{array}{cc}
  \Sigma_{1,1} & \Sigma_{1,2} \\
  \Sigma_{1,2}^\top & \Sigma_{2,2}
\end{array}
\right],\Sigma_{1,1}\in \mathbb{S}^{p-1}_{++}.
\end{equation}
\end{property}

\begin{property}\label{property_4}
Some special cases of the GS model are listed in Table \ref{specialcase_table}.
\begin{table}[htbp]
\begin{center}
\caption{Special cases of the GS model with corresponding parameters}\label{specialcase_table}
\begin{tabular}{cc}
\toprule
Special case& Parameter value\\
\midrule
bursty IN& $\rho=0$ and $\tilde{\Sigma}\neq\mathbf{I}_p$\\
WGN& $p=1$ or $\rho=1$\\
white IN& $\rho=0$ and $p=1$\\
white mixed noise& $p=1$\\
\bottomrule
\end{tabular}
\end{center}
\end{table}
\end{property}

\section{Demodulation of PSK signals under bursty mixed noise}\label{application}
In this section, we consider the demodulation of PSK signals under the bursty mixed noise. For other linear modulation schemes, the corresponding demodulation schemes can be achieved via similar procedures.

\subsection{Received signal model}\label{subsection_Received_signal_model}
To mitigate the negative influence of noise clusters, the relationship among multiple neighboring noise samples should be considered. Therefore, the sampling rate usually needs to be larger than the symbol rate, i.e., there are multiple samples obtained during a symbol period. Let $f_s$ and $T$ be the sampling rate and the symbol period, respectively. We denote the oversampling rate by $K=f_sT$ in the sequel. The received baseband signal of PSK modulation within a symbol period can be expressed as
\begin{equation}\label{received_signal_model}
y_i(j)=x_i+n_i(j),j=1,2,\cdots,K,
\end{equation}
where $x_i\in\Omega(M)=\{\omega_M^k=A\exp(j2\pi k/M),k=1,\cdots,M\}$ represents the $i$-th M-array PSK (M-PSK) symbol and $A$ is the signal amplitude. $\Omega(M)$ is the feasible set of M-PSK modulation signals. $y_i(j)$ and $n_i(j)$ are the $j$-th samples of the received signal and noise during the $i$-th symbol period. Note that all of the samples are complex and we will denote the in-phase and quadrature transmitted symbol by $x_i^I$ and $x_i^Q$, respectively. For national simplicity, we neglect the time index for $x_i$ since it remains unchanged within a symbol period.

\subsection{Demodulation Algorithm}\label{Demodulation_of_PSK}
To derive the optimal demodulation algorithm in the bursty mixed noise, the complex GS model is needed. Without loss of generality, we assume that the in-phase and quadrature noise components are mutually independent. \eqref{received_signal_model} is equivalent to
\begin{align}\label{received_signal_model_vector}
\mathbf{y}_i=x_i\mathbf{e}_K+\mathbf{n}_i
\end{align}
where $\mathbf{y}_i=[y_i(1),\cdots,y_i(K)]^\top$, $\mathbf{n}_i=[n_i(1),\cdots,n_i(K)]^\top$, $\mathbf{n}_i=\mathbf{n}_i^I+j\mathbf{n}_i^Q$ and $\mathbf{e}_K\in\mathbb{R}^{K\times1}$ is a vector with all entries one. Then, the PDF of complex GS model can be given as
\begin{align}\label{complex_GS_model}
f_C\left(\mathbf{n}^I,\mathbf{n}^Q\right)=f_M\left(\mathbf{n}^I\right)f_M\left(\mathbf{n}^Q\right)
\end{align}
where $f_M(\cdot)$ is provided in \eqref{bursty_mixed_noise_model}. Meanwhile, the conditional PDF is
\begin{align}\label{complex_conditional_PDF}
f_C\left(n^I,n^Q|\mathbf{n}_1^I,\mathbf{n}_1^Q\right)=f_M\left(n^I|\mathbf{n}_1^I\right)f_M\left(n^Q|\mathbf{n}_1^Q\right)
\end{align}
where $\mathbf{n}^I=[(\mathbf{n}_1^I)^\top,n^I]^\top$ and $\mathbf{n}^Q=[(\mathbf{n}_1^Q)^\top,n^Q]^\top$. Our goal is to maximize the following log-likelihood function,
\begin{equation}\label{demodulation_problem_formulation}
\hat{x}_i=\mathop{\arg\max}\limits_{x_i\in\Omega(M)}{\log\left[f_J(\mathbf{y}_i-x_i\mathbf{e}_K)\right]}.
\end{equation}
where $f_J(\cdot)$ represents the joint probability density function. Note that we have $f_J(\mathbf{y}_i-x_i)=f_C(\mathbf{y}_i-x_i)$ if $K\leq p$. Therefore, the explicit expression of $f_J(\mathbf{y}_i-x_i\mathbf{e}_K)$ with $K>p$ has to be determined by the probability chain rule as following,
\begin{align}\label{likelihood_function_1}
f_J(\mathbf{y}_i-x_i\mathbf{e}_K)=&f_J(y_i(1)-x_i)\nonumber\\
&\times\prod_{j=2}^{K}f_J\left(y_i(j)-x_i|\mathbf{y}_i^{1,j-1}-x_i\mathbf{e}_{j-1}\right),
\end{align}
where $\mathbf{y}_i^{j,k}=[y_i(j),\cdots,y_i(k)]^\top$. It follows the Markov property that the $j$-th sample is only relevant to the previous $p-1$ samples. Consequently,
\begin{align}\label{likelihood_function}
f_J(\mathbf{y}_i-x_i\mathbf{e}_{K})=&f_C\left(\mathbf{y}_i^{1,p}-x_i\mathbf{e}_{p}\right)\prod_{j=p+1}^{K}f_C\Big(y_i(j)-x_i|\nonumber\\
&\mathbf{y}_i^{j-p+1,j-1}-x_i\mathbf{e}_{p-1}\Big).
\end{align}

Furthermore, with the help of \eqref{complex_GS_model} and \eqref{complex_conditional_PDF},
\begin{align}\label{final_expression_joint_PDF}
&f_J(\mathbf{y}_i-x_i\mathbf{e}_{K})\nonumber\\
=&f_M\left((\mathbf{y}_i^I)^{1,p}-(x_i^I)\mathbf{e}_{p}\right)f_M\left((\mathbf{y}_i^Q)^{1,p}-(x_i^Q)\mathbf{e}_{p}\right)\nonumber\\
&\times\prod_{j=p+1}^{K}\Big[f_M\Big(y_i^I(j)-x_i^I|(\mathbf{y}_i^I)^{j-p+1,j-1}-x_i^I\mathbf{e}_{p-1}\Big)\nonumber\\
&\times f_M\Big(y_i^Q(j)-x_i^Q|(\mathbf{y}_i^Q)^{j-p+1,j-1}-x_i^Q\mathbf{e}_{p-1}\Big)\Big].
\end{align}

Then, by combining \eqref{bursty_mixed_noise_model} and Property \ref{property_2}, \eqref{likelihood_function} can be calculated. \eqref{final_expression_joint_PDF} admits analytic expression and the computational complexity is not considerable since there is no numerical integration.

\subsection{Special cases}\label{subsection_special_case}
To facilitate practical applications, we subsequently discuss the simplified forms of \eqref{final_expression_joint_PDF} for several special cases.
\subsubsection{Special noise cases}
When the bursty mixed noise degenerates to WGN in accordance with Property \ref{property_4}, we have
\begin{align}
\log[f_J(\mathbf{y}_i-x_i\mathbf{e}_{K})]\propto&\Vert\Delta(\mathbf{u}^I_i)^{1,p}\Vert^2+\Vert\Delta(\mathbf{u}^Q_i)^{1,p}\Vert^2+\sum_{j=p+1}^{K}\nonumber\\
&\left(|\Delta u^I_i(j)|^2+|\Delta u^Q_i(j)|^2\right),
\end{align}
where we denote that $\Delta(\mathbf{u}^I_i)^{p,q}\triangleq(\mathbf{y}_i^I)^{p,q}-(x_i^I)\mathbf{e}_{p-q+1}$ and $\Delta u^I_i(j)\triangleq y_i^I(j)-x_i^I$ for simplicity, and the quadrature signal component follows the similar notations. It can be seen that the ML metric becomes the Euclidean distance. Meanwhile, if the background gaussian noise of bursty mixed noise can be neglected, the likelihood function is
\begin{align}
&\log[f_J(\mathbf{y}_i-x_i\mathbf{e}_{K})]\nonumber\\
\propto&\log[\alpha+\Vert\Sigma^{-1/2}\Delta(\mathbf{u}^I_i)^{1,p}\Vert^2]+\log[\alpha+\Vert\Sigma^{-1/2}\Delta(\mathbf{u}^Q_i)^{1,p}\Vert^2]\nonumber\\
&+\frac{\alpha+p-1}{\alpha+p}\left(\Delta\mathbf{v}_{i}^{I}+\Delta\mathbf{v}_{i}^{Q}\right),
\end{align}
where
\begin{align}
\Delta\mathbf{v}_{i}^{l}=&\sum_{j=p+1}^{K}\bigg\{\log\bigg[\alpha+\bigg(\Vert\Sigma_{1,1}^{-1/2}\Delta(\mathbf{u}^l_i)^{j-p+1,j-1}\Vert^2\nonumber\\
&+\frac{(\Delta u^l_i(j)-\mu(j))^2}{\sigma}\bigg)\bigg]-\log\Big[\alpha+\Vert\Sigma_{1,1}^{-1/2}\nonumber\\
&\times\Delta(\mathbf{u}^l_i)^{j-p+1,j-1}\Vert^2\Big]\bigg\},l=I,Q,
\end{align}
and $\mu(j)=[\Delta(\mathbf{u}^l_i)^{j-p+1,j-1}]^\top\Sigma_{1,1}^{-1}\Sigma_{1,2}$. In fact, the ML demodulation under the pure bursty IN is similar to the myriad detector \cite{paper20}.
For the other special noise cases given in Property \ref{property_4}, the corresponding simplified demodulation algorithms can be similarly analyzed.

\subsubsection{Special modulation schemes}

We further consider the demodulation algorithms for BPSK and QPSK. By considering $\Omega(2)=\{-1,1\}$, the demodulation of BPSK can be formulated as follows,
\begin{equation}\label{simplification_BPSK}
\hat{x}_i=\mathop{\arg\max}\limits_{x_i\in\Omega(2)}{\log\left[f_M(\mathbf{y}_i-x_i\mathbf{e}_K)\right]}.
\end{equation}
Notice that the $x_i$ herein is a real number.

For QPSK signals, the in-phase and quadrature signals can be considered as two parallel BPSK modulated signals and detected separately, i.e.,
\begin{align}\label{simplification_QPSK}
&\Re\{\hat{x}_i\}=\mathop{\arg\max}\limits_{x_i\in\Omega(4)}{\log\left[f_M(\Re\{\mathbf{y}_i\}-\Re\{x_i\}\mathbf{e}_K)\right]},\nonumber\\
&\Im\{\hat{x}_i\}=\mathop{\arg\max}\limits_{x_i\in\Omega(4)}{\log\left[f_M(\Im\{\mathbf{y}_i\}-\Im\{x_i\}\mathbf{e}_K)\right]},
\end{align}
where $\Re\{\cdot\}$ and $\Im\{\cdot\}$ denote the real and imaginary part, respectively. In this way, the computational complexity of likelihood function can be considerably reduced.
%

\subsection{Performance analysis}
Compared with the BER analysis with respect to white noise \cite{book3},  it is much more challenging when the noise has memory, i.e., color noise, due to the resulting high-dimensional integral. Therefore, by combining \eqref{simplification_BPSK} and \eqref{simplification_QPSK}, we derive the asymptotic BER of BPSK and QPSK with analytical expressions. As the complex GS model is rather complicated, we analyze the error probability bounds of M-PSK for $M>4$. By considering different complexity, we derive the BER for the case of $p=1$ and $p\geq2$, respectively. Meanwhile, as BPSK and QPSK have the same theoretical error probability, we first consider the BER of BPSK.

Without loss of generality, we consider the case of $K=p$, which can be satisfied by setting $f_s=p/T$. In this way, we have $f_J(\mathbf{n}_i)=f_M(\mathbf{n}_i)$ based on \eqref{simplification_BPSK}. The subscript $i$ is omitted in what follows for brevity. Meanwhile, we assume that $X$ usually follows uniform distribution, i.e., $P(X=A)=P(X=-A)=1/2$. The BER of BPSK can then be expressed as
\begin{align}\label{BER_original_expression}
P_e=&\frac{1}{2}P(f_M(\mathbf{y}+A\mathbf{e}_p)>f_M(\mathbf{y}-A\mathbf{e}_p)|x=A)\nonumber\\
&+\frac{1}{2}P(f_M(\mathbf{y}+A\mathbf{e}_p)<f_M(\mathbf{y}-A\mathbf{e}_p)|x=-A).
\end{align}

Based on the symmetry of $f_M(\cdot)$, \eqref{BER_original_expression} equals to
\begin{align}\label{BER_original_expression_1}
P_e=P(f_M(\mathbf{n}+2A\mathbf{e}_p)>f_M(\mathbf{n})).
\end{align}

Subsequently, we analyze \eqref{BER_original_expression_1} by considering two cases. We first derive the BER under the special case $p=1$, for which the noise has no bursty property and \eqref{bursty_mixed_noise_model} degenerates to the univariate distribution. For the case $p\geq2$, \eqref{BER_original_expression_1} is a high-dimensional integral. To the best of our knowledge, there is little analytical BER expression with respect to a high-dimensional distribution due to its complexity. The most popular method is to leverage the central limit theorem (CLT) to approximate the sum of RVs as a Gaussian RV. However, the CLT does not hold due to the correlation between noise samples in our considering problem. Moreover, the CLT method still needs numerical calculation. In this subsection, we derive the asymptotic BER with closed-form expression.

\subsubsection{Case of $p=1$}
In this case, \eqref{bursty_mixed_noise_model} becomes
\begin{align}\label{one_dimension}
f_M(n,p=1)=&\frac{\rho}{2\gamma_s\sqrt{\pi}}\exp\left(-\frac{n^2}{4\gamma_g^2}\right)+(1-\rho)\nonumber\\
&\times\frac{\Gamma((\alpha+1)/2)}{\sqrt{2\alpha\pi}\gamma_s\Gamma(\alpha/2)}\left(1+\frac{n^2}{2\alpha\gamma_s^2}\right)^{-\frac{\alpha+1}{2}}.
\end{align}

Then, \eqref{BER_original_expression_1} becomes
\begin{align}\label{BER_P_1}
P_e=&P\left(f_M(n+2A,p=1)>f_M(n,p=1)\right) \nonumber\\
=&\int_{A}^{+\infty}f_M(n,p=1)dn \nonumber\\
=&\frac{\rho}{2\gamma_s\sqrt{\pi}}\int_{A}^{+\infty}\exp\left(-\frac{n^2}{4\gamma_g^2}\right)dn+(1-\rho)\nonumber\\
&\times\frac{\Gamma((\alpha+1)/2)}{\sqrt{2\alpha\pi}\gamma_s\Gamma(\alpha/2)}\int_{A}^{+\infty}\left(1+\frac{n^2}{2\alpha\gamma_s^2}\right)^{-\frac{\alpha+1}{2}}dn.
\end{align}

Before proceeding, we first introduce the following integral identity \cite{book4},
\begin{equation}\label{integral_identity}
\int \frac{r^a}{(b+r^2)^c}=\frac{r^{a+1}}{b(a+1)}{_{2}F_1}\left(1,\frac{a+1}{c};\frac{a+c+1}{c};-\frac{r^c}{b}\right),
\end{equation}
where ${_{2}F_1}(\cdot,\cdot;\cdot;\cdot)$ is the Gaussian hypergeometric function. Then, \eqref{BER_P_1} can be simplifies as follows,
\begin{align}
P_e=&\rho Q\left(\frac{\sqrt{2}A}{2\gamma_g}\right)+\frac{(1-\rho)\Gamma((\alpha+1)/2)}{\Gamma(\alpha/2)\gamma_s\sqrt{2\alpha\pi}}\bigg\{\lim\limits_{z\rightarrow+\infty}\bigg[z{_{2}F_1}\bigg(\frac{1}{2}\nonumber\\
&,\frac{\alpha+1}{2};\frac{3}{2};\frac{-z^2}{2\alpha\gamma_s^2}\bigg)\bigg]-A{_{2}F_1}\bigg(\frac{1}{2},\frac{\alpha+1}{2};\frac{3}{2};\frac{-A^2}{2\alpha\gamma_s^2}\bigg)\bigg\},
\end{align}
where
\begin{equation}
Q(z)=\frac{1}{\sqrt{2\pi}}\int_{z}^{+\infty}\exp\left(-\frac{t^2}{2}\right)dt.
\end{equation}

Considering the following linear transform equation of the Gaussian hypergeometric function \cite{book4}
\begin{align}\label{linear_trans}
&{_{2}F_1}\left(a,b;c;z\right)\nonumber\\
=&\frac{\Gamma(c)\Gamma(b-a)}{\Gamma(b)\Gamma(c-a)}(-z)^{-a}{_{2}F_1}\left(a,a+1-c;a+1-b;z^{-1}\right)\nonumber\\
&+\frac{\Gamma(c)\Gamma(a-b)}{\Gamma(a)\Gamma(c-b)}(-z)^{-b}{_{2}F_1}\left(b,b+1-c;b+1-a;z^{-1}\right),
\end{align}
we have
\begin{align}
\lim\limits_{z\rightarrow+\infty}\bigg[z{_{2}F_1}\bigg(\frac{1}{2},\frac{\alpha+1}{2};\frac{3}{2};\frac{-z^2}{2\alpha\gamma_s^2}\bigg)\bigg]=\frac{\sqrt{2\alpha}\gamma_s\Gamma(\frac{3}{2})\Gamma(\frac{\alpha}{2})}{\Gamma(\frac{\alpha+1}{2})}.
\end{align}

Then, we can get the BER expression of BPSK as follows,
\begin{align}
P_e=&\rho Q\left(\frac{\sqrt{2}A}{2\gamma_g}\right)+\frac{(1-\rho)\Gamma((\alpha+1)/2)}{\Gamma(\alpha/2)\gamma_s\sqrt{2\alpha\pi}}\bigg\{\frac{\sqrt{2\alpha}\gamma_s\Gamma(\frac{3}{2})\Gamma(\frac{\alpha}{2})}{\Gamma(\frac{\alpha+1}{2})}\nonumber\\
&-A{_{2}F_1}\bigg(\frac{1}{2},\frac{\alpha+1}{2};\frac{3}{2};\frac{-A^2}{2\alpha\gamma_s^2}\bigg)\bigg\}.
\end{align}

$\mathbf{Remarks}$: The logarithm of the first term, i.e., $\log[Q\left(\sqrt{2}A/2\gamma_g\right)]$, has a quadratic slope which corresponds to the decreasing rate of the BER versus signal-to-noise ratio (SNR) under AWGN channel. It decreases very quickly as the signal amplitude continues to increase. In contrast, the logarithm of the last term $\log\left[A{_{2}F_1}\left(\frac{1}{2},\frac{\alpha+1}{2};\frac{3}{2};\frac{-A^2}{2\alpha\gamma_s^2}\right)\right]$ decays linearly with respect to $A$. Consequently, the BER versus SNR will decrease with the quadratic slope in the low SNR region, and decrease nearly linearly with the increase of SNR.

\subsubsection{Case of $p\geq2$}
For the case of $p\geq2$, \eqref{BER_original_expression_1} is a high-dimensional integral, i.e.,
\begin{equation}
P_e=\int_{\Omega}f_J(\mathbf{n})d\mathbf{n}=\int_{\Omega}f_M(\mathbf{n})d\mathbf{n},
\end{equation}
where $\Omega=\{\mathbf{n}:f_M(\mathbf{n}+2A\mathbf{e}_p)>f_M(\mathbf{n})\}$ is the integral region. When $\tilde{\Sigma}=\mathbf{I}_p$, the region $\Omega$ becomes
\begin{align}\label{region_identical_matrix}
\Omega_1=&\left\{\mathbf{n}:f_M(\mathbf{n}+2A\mathbf{e}_p)>f_M(\mathbf{n}),\tilde{\Sigma}=\mathbf{I}_p\right\}\nonumber\\
=&\left\{\mathbf{n}:\Vert\mathbf{n}+2A\mathbf{e}_p\Vert^2<\Vert\mathbf{n}\Vert^2\right\}\nonumber\\
=&\left\{\mathbf{n}:\mathbf{n}^\top\mathbf{e}_p<-pA\right\}.
\end{align}

However, the integral region $\Omega$ is not trivial and rather difficult to express in general. Then, we will analyze the asymptotic expression. We first derive simple lower and upper bounds. Let
\begin{equation}
\Omega_A=\left\{\mathbf{n}:f_M(\mathbf{n}+2A\mathbf{e}_p)>f_M(\mathbf{n}),\rho=1\right\},
\end{equation}
\begin{equation}
\Omega_B=\left\{\mathbf{n}:f_M(\mathbf{n}+2A\mathbf{e}_p)>f_M(\mathbf{n}),\rho=0\right\}.
\end{equation}

It can be observed that $\Omega_A=\Omega_1$. Meanwhile,
\begin{align}\label{Omega_B}
\Omega_B=&\left\{\mathbf{n}:f_M(\mathbf{n}+2A\mathbf{e}_p)>f_M(\mathbf{n}),\rho=0\right\} \nonumber\\
=&\left\{\mathbf{n}:\Vert\Sigma^{-1/2}(\mathbf{n}+2A\mathbf{e}_p)\Vert^2<\Vert\Sigma^{-1/2}\mathbf{n}\Vert^2\right\}\nonumber\\
=&\left\{\mathbf{n}:(\mathbf{n}+2A\mathbf{e}_p)^\top\Sigma^{-1}(\mathbf{n}+2A\mathbf{e}_p)<\mathbf{n}^\top\Sigma^{-1}\mathbf{n}\right\}\nonumber\\
=&\left\{\mathbf{n}:\mathbf{n}^\top\Sigma^{-1}\mathbf{e}_p<-A\mathbf{e}_p^\top\Sigma^{-1}\mathbf{e}_p\right\},
\end{align}

The integral region $\Omega_B$ will become smaller with the increase of $A$. Both $\Omega_A$ and $\Omega_B$ represent the $p$-dimensional half-spaces and $\Omega_B$ is an affine transformation of $\Omega_A$. Moreover, we have$(\Omega_A\cap\Omega_B)\subset\Omega$ and $\Omega \subset (\Omega_A\cup\Omega_B)$. Therefore,
\begin{align}\label{naive_bounds}
\int_{\Omega_A\cap\Omega_B}f_M(\mathbf{n})d\mathbf{n}\leq P_e(\tilde{\Sigma}\neq\mathbf{I}_p)\leq \int_{\Omega_A\cup\Omega_B}f_M(\mathbf{n})d\mathbf{n}.
\end{align}

The tightness of these bounds is determined by $\tilde{\Sigma}$. The bounds converge to the theoretical BER if $\Sigma^{-1}\mathbf{e}_p=k\mathbf{e}_p$ where $k$ is an arbitrary constant, which is always satisfied when $p=2$. Consequently,
\begin{equation}
P_e(\tilde{\Sigma}\neq\mathbf{I}_p,p=2)=\int_{\Omega_1}f_M(\mathbf{n})d\mathbf{n}.
\end{equation}

On the other hand, the integration of the exponential term declines very quickly when the integral region is far from the origin, i.e., $A$ is larger. Then, the exponential term can be neglected, and the upper bound will be rather tight and converge to the theoretical BER for large SNR. Therefore, the asymptotic expression of theoretical BER can be given as
\begin{align}\label{lower_bound_inequality_1}
P_e(\tilde{\Sigma}\neq\mathbf{I}_p)\approx &\int_{\Omega_B}f_M(\mathbf{n})d\mathbf{n} \nonumber\\
\approx&\int_{\Omega_B}\frac{(1-\rho)k_2}{\left(1+\Vert\Sigma^{-1/2}\mathbf{n}\Vert^2/\alpha\right)^\frac{\alpha+p}{2}}d\mathbf{n}\triangleq \hat{P}_e.
\end{align}

Finally, the asymptotic BER is summarized as the following theorem.
\begin{theorem}\label{theorem_3}
Let the memory order $p\geq2$. It follows the PDF \eqref{bursty_mixed_noise_model} that the asymptotic BER $P_e$ of BPSK and QPSK modulation with ML demodulation is
\begin{align}\label{asymptotic_BER}
\hat{P}_e=&\frac{1-\rho}{2}\bigg[1-\frac{P_0\Gamma(\frac{\alpha+1}{2})}{\lambda\sqrt{\alpha}\Gamma(\frac{3}{2})\Gamma(\frac{\alpha}{2})}{_{2}F_1}\bigg(\frac{1}{2},\frac{\alpha+1}{2};\frac{3}{2};-\frac{P_0^2}{\alpha\lambda^{2}}\bigg)\bigg],
\end{align}
where $P_0=A\mathbf{e}_p^\top\Sigma^{-1}\mathbf{e}_p$ and $\lambda=\Vert(\Sigma^{-1}\mathbf{e}_p)^\top\Sigma^{-1/2}_{inv}\Vert$.
\end{theorem}
\begin{IEEEproof}
The proof is relegated to Appendix \ref{appendix_theorem_3}.
\end{IEEEproof}

In fact, the theorem \ref{theorem_3} is rather general. We can obtain accurate theoretical BER expressions for some special cases via the similar procedures. For example, the BER expressions for the following 3 special cases can be obtained straightforwardly.
\begin{corollary}\label{corollary_1}
Let $p\geq2,\rho=1$, the accurate theoretical BER is
\begin{align}\label{corollary_1_ber}
P_e(p\geq2,\rho=1)=Q\left(\frac{A\sqrt{2p}}{2\gamma_g}\right).
\end{align}
\end{corollary}

Obviously, \eqref{corollary_1_ber} is the same as the BER of M-PSK modulation in the AWGN channel.
\begin{corollary}\label{corollary_2}
Let $p\geq2,\tilde{\Sigma}=\mathbf{I}_p$, the accurate theoretical BER is
\begin{align}
&P_e(p\geq2,\tilde{\Sigma}=\mathbf{I}_p)\nonumber\\
=&\frac{1-\rho}{2}\bigg[1-\frac{\sqrt{p}A\Gamma(\frac{\alpha+1}{2})}{\sqrt{2\alpha}\gamma_s\Gamma(\frac{3}{2})\Gamma(\frac{\alpha}{2})}{_{2}F_1}\bigg(\frac{1}{2},\frac{\alpha+1}{2};\frac{3}{2};-\frac{pA^2}{2\alpha\gamma_s^2}\bigg)\bigg].
\end{align}
\end{corollary}

The region $\Omega_B$ is equivalent to $\Omega_A$ if $\tilde{\Sigma}=\mathbf{I}_p$. According to the theorem \ref{theorem_3} and corollary \ref{corollary_1}, we obtain the corollary \ref{corollary_2} by setting $\tilde{\Sigma}=\mathbf{I}_p$.

\begin{corollary}\label{corollary_3}
Let $p\geq2,\rho=0$, the accurate theoretical BER is the same as \eqref{asymptotic_BER}.
\end{corollary}

\subsubsection{Case of $M>4$}
To reach optimal demodulation performance when $M>4$, the complex GS distribution has to be leveraged. Therefore, it is quite difficult to derive the analytical and accurate BER expression due to the complicated integral region and noise model. Thus, we convert to investigate the error probability bounds. Note that both the symbol error rate (SER) and BER can measure the demodulation performance. By default, the SER of M-PSK modulation ($M>4$) is considered in what follows.

We define the signal index set of M-PSK to be $\Lambda_M=\{1,2,\cdots,M\}$. According to \eqref{complex_conditional_PDF} and \eqref{likelihood_function}, the error probability of M-PSK signal given the input symbol $\omega_M^i\in \Omega(M)$, $i\in \Lambda_M$ equals
\begin{align}\label{SER_expression_MPSK}
&P_e^M(\omega_M^i)\nonumber\\
=&\frac{1}{M}\sum_{k\in\Lambda_M}P\left(f_C\left(\mathbf{y}^{1,p}-\omega_M^i\mathbf{e}_{p}\right)<f_C\left(\mathbf{y}^{1,p}-\omega_M^k\mathbf{e}_{p}\right)\right)\nonumber\\
=&\frac{1}{M}\sum_{k\in\Lambda_M}P\Big(f_M((\mathbf{y}^I)^{1,p}-(\omega_M^i)^I\mathbf{e}_{p})f_M((\mathbf{y}^Q)^{1,p}-(\omega_M^i)^Q\nonumber\\
&\times\mathbf{e}_{p})<f_M((\mathbf{y}^I)^{1,p}-(\omega_M^k)^I\mathbf{e}_{p})f_M((\mathbf{y}^Q)^{1,p}-(\omega_M^k)^Q\mathbf{e}_{p})\Big)\nonumber\\
\triangleq&\frac{1}{M}\sum_{k\in\Lambda_M}P_e^M(\omega_M^i,k).
\end{align}

For the sake of simplicity, we define
\begin{align}
\mathcal{A}^l_{i,k}\triangleq&\big\{\mathbf{n}^l:f_M((\mathbf{y}^l)^{1,p}-(\omega_M^i)^l\mathbf{e}_{p})<f_M((\mathbf{y}^l)^{1,p}\nonumber\\
&-(\omega_M^k)^l\mathbf{e}_{p}),l=I,Q\big\}.
\end{align}

The possible lower and upper bound of $P_e^M(\omega_M^i,k)$ can be represented as $P_e^M(\omega_M^i,k)\geq P(\mathcal{A}^I_{i,k})P(\mathcal{A}^Q_{i,k})$ and $P_e^M(\omega_M^i,k)\leq P(\mathcal{A}^I_{i,k})+P(\mathcal{A}^Q_{i,k})-P(\mathcal{A}^I_{i,k})P(\mathcal{A}^Q_{i,k})$, respectively. Unfortunately, these bounds are not tight enough for practical applications. Note that the $P_e^M(\omega_M^i,k)$ can be further simplified for some special $\omega_M^k$. For instance, $\mathcal{A}^l_{i,k}$ is a null set when $i=k$ and therefore, $P_e^M(\omega_M^i,i)=0$. Moreover, $P_e^M(\omega_M^i,k)=P(\mathcal{A}^l_{i,k})$ if $(\omega_M^i)^{l^c}=(\omega_M^k)^{l^c}$ where $l^c$ represents the complement element of $l$, e.g., $l^c=Q$ if $l=I$. In this case, the constellation points $\omega_M^i$ and $\omega_M^k$ are symmetric about a coordinate axis. In fact, if they are symmetric about the origin, i.e., $|(\omega_M^i)^I-(\omega_M^k)^I|=|(\omega_M^i)^Q-(\omega_M^k)^Q|$, it can be shown that $P_e^M(\omega_M^i,k)=P(\mathcal{A}^l_{i,k})$. The close-form expressions of $P(\mathcal{A}^l_{i,k})$ can be calculated based on Theorem \ref{theorem_3} with similar procedures. By replacing $P_0$ with $P_0^{i,k,l}$ in \eqref{asymptotic_BER}, we can conclude that
\begin{align}\label{expression_A_ikl}
&P\left(\mathcal{A}^l_{i,k}\right)\nonumber\\
\approx&\frac{1-\rho}{2}\bigg[1-\frac{P_0^{i,k,l}\Gamma(\frac{\alpha+1}{2})}{\lambda\sqrt{\alpha}\Gamma(\frac{3}{2})\Gamma(\frac{\alpha}{2})}{_{2}F_1}\bigg(\frac{1}{2},\frac{\alpha+1}{2};\frac{3}{2};-\frac{(P_0^{i,k,l})^2}{\alpha\lambda^{2}}\bigg)\bigg]
\end{align}
where
\begin{align}
P_0^{i,k,l}=\frac{|(\omega_M^i)^l-(\omega_M^k)^l|\mathbf{e}_p^\top\Sigma^{-1}\mathbf{e}_p}{2}
\end{align}

Except for the aforementioned special cases, $P_e^M(\omega_M^i,k)$ could not degrade to an analytical and asymptotic expression. However, it can be easily shown that all of the $\mathcal{A}^l_{i,k}$ with fixed $l$ constructs a monotone class \cite{book7} and the size of a set is measured by the $|(\omega_M^i)^l-(\omega_M^k)^l|$ which is related to the subscripts. In other words, we have $\mathcal{A}^l_{i,k_1}\subseteq\mathcal{A}^l_{i,k_2}$ and $P(\mathcal{A}^l_{i,k_1})>P(\mathcal{A}^l_{i,k_2})$ if $|(\omega_M^i)^l-(\omega_M^{k_1})^l|<|(\omega_M^i)^l-(\omega_M^{k_2})^l|$, which means that the larger the Euclidean distance of two constellation points is, the smaller the error probability is. Therefore, assuming that $|(\omega_M^i)^I-(\omega_M^{k})^I|<|(\omega_M^i)^{Q}-(\omega_M^{k})^{Q}|$, the general lower and upper bounds of the $P_e^M(\omega_M^i,k)$ can be obtained as follows,
\begin{align}
P_e^M(\omega_M^i,k)\leq P\left(\mathcal{A}^I_{i,k}\right),
\end{align}
\begin{align}
P_e^M(\omega_M^i,k)\geq P\left(\mathcal{A}^Q_{i,k}\right).
\end{align}

Finally, the SER lower and upper bound of M-PSK ($M>4$) signals can be given as
\begin{align}\label{upper_bound_MPSK}
P_e^M\leq\frac{1}{M^2}\sum_{i\in\Lambda_M}\sum_{k\in\Lambda_M}\mathcal{T}_U(\omega_M^i,k)
\end{align}
and
\begin{align}\label{lower_bound_MPSK}
P_e^M\geq\frac{1}{M^2}\sum_{i\in\Lambda_M}\sum_{k\in\Lambda_M}\mathcal{T}_L(\omega_M^i,k)
\end{align}
where
\begin{equation}\label{upper_SER_function_MPSK}
\mathcal{T}_U(\omega_M^i,k)=\left\{
\begin{aligned}
&0,i=k\\
&P\left(\mathcal{A}^l_{i,k}\right),(\omega_M^i)^{l^c}=(\omega_M^k)^{l^c}\text{ or }|(\omega_M^i)^l \\
&-(\omega_M^{k})^l|\leq|(\omega_M^i)^{l^c}-(\omega_M^{k})^{l^c}|
\end{aligned}
\right.
\end{equation}

\begin{equation}\label{lower_SER_function_MPSK}
\mathcal{T}_L(\omega_M^i,k)=\left\{
\begin{aligned}
&0,i=k\\
&P\left(\mathcal{A}^l_{i,k}\right),(\omega_M^i)^{l^c}=(\omega_M^k)^{l^c}\text{ or }|(\omega_M^i)^l \\
&-(\omega_M^{k})^l|\geq|(\omega_M^i)^{l^c}-(\omega_M^{k})^{l^c}|
\end{aligned}
\right.
\end{equation}

$\mathbf{Remarks}$: Specifically, we have $\mathcal{T}_U(\omega_M^i,k)\geq P_e^M(\omega_M^i,k)$ in general and the equality can be achieved except for the case $|(\omega_M^i)^l-(\omega_M^{k})^l|<|(\omega_M^i)^{l^c}-(\omega_M^{k})^{l^c}|$. In addition, both of \eqref{upper_bound_MPSK} and \eqref{lower_bound_MPSK} admit closed forms with the aid of theorem \ref{theorem_3}. The bounds shown in \eqref{upper_bound_MPSK} and \eqref{lower_bound_MPSK} will degenerate to theorem \ref{theorem_3} for $M=2$ or $4$. Furthermore, they are also asymptotic since theorem \ref{theorem_3} is derived with respect to the noise model approximation for the high SNR region. It is worth noting that the SER of M-QAM signals can be similarly analyzed.

\section{Demodulation of MSK signals under bursty mixed noise}\label{section_demodulation_MSK}
As a typical nonlinear modulation scheme, MSK has been widely used in non-Guassian IN noise channels \cite{book2,paper20}. In this section, we consider the demodulation algorithm of MSK signals and analyze the corresponding theoretical performance to get the closed-form lower and upper bound of error probability.

\subsection{Received signal model}
Considering the passband signal model, the received MSK signal of the $i$-th symbol can be expressed as follows,
\begin{align}\label{MSK_received_signal_model}
\mathbf{y}_i=\mathbf{s}_{i,\theta}+\mathbf{n}_i,i=1,2,\cdots
\end{align}
where $\mathbf{y}_i=[y_i(1),\cdots,y_i(K)]^{\top}$ and $K$ is the oversampling rate. $\mathbf{n}_i$ is the bursty mixed noise vector and $\mathbf{s}_{i,\theta}\subset \mathbb{R}^K$ is the MSK modulated signal vector with element $s_{i,\theta}(j)$ given as \cite{book3}
\begin{align}\label{MSK_signal_model}
s_{i,\theta}(j)=A\cos\left[\left(2\pi f_c+\frac{d_i\pi}{2T}\right)\frac{j}{f_s}+\theta_i\right],j=1,2,\cdots,K,
\end{align}
where $A$ and $T$ denote the amplitude and the symbol period of the MSK signals, respectively. $f_c$ is the carrier frequency and $f_s$ is the sampling rate. Note that the oversampling rate $K$ equals $f_sT$. $\theta_i$ represents the original phase of the $i$-th symbol and $d_i$ is the $i$-th transmitted bipolar symbol. For MSK modulated signals, $d_i\in\{-1,+1\}$ and $\theta_i\in\Omega_{\theta}=\{0,\pi/2,\pi,3\pi/2\}$ and therefore, the feasible set of $\mathbf{s}_{i,\theta}$ has 8 elements.

\subsection{Demodulation of MSK signal based on Viterbi algorithm}
VA has been widely used to demodulate the MSK signals due to its low complexity and near-optimal performance \cite{book3}. For VA based MSK demodulation algorithm, the key point is to calculate the branch metric. It follows \eqref{bursty_mixed_noise_model} and the methods presented in subsection \ref{Demodulation_of_PSK}, we can define the ML metric as
\begin{align}\label{Viterbi_ML_metric}
M_{ML}&=\log\left[f_J\left(\mathbf{y}_i-\mathbf{s}_{i,\theta}\right)\right]\nonumber\\
=&\log f_M\left(\mathbf{y}_i^{1,p}-\mathbf{s}_{i,\theta}^{1,p}\right)+\sum_{j=p+1}^{K}\log f_M\Big(y_i(j)-s_{i,\theta}(j)|\nonumber\\
&\mathbf{y}_i^{j-p+1,j-1}-\mathbf{s}_{i,\theta}^{j-p+1,j-1}\Big).
\end{align}

In this way, we can construct the state transition and determine the transmitted symbol by searching the path with the largest branch metric. Notice that the ML metric can also be leveraged for CPM signal with higher modulation order by slight modifications. Meanwhile, the special case analysis in subsection \ref{subsection_special_case} is still applicable for the nonlinear modulation schemes.

\subsection{Performance bounds}\label{subsection_performance_bound_MSK}
In this subsection, the theoretical BER performance of MSK signals demodulated by VA is further analyzed. For notational simplicity, the time index is omitted in what follows. The accurate theoretical BER is very challenging to derive since each symbol is related to all the previous states, which results in an enormous amount of error events. Therefore, some approximations have to be applied. Here, we propose a lower bound and upper bound of the BER by considering the error event with the largest probability.

\begin{figure}[htbp]
\centering
\includegraphics[width=3.3in]{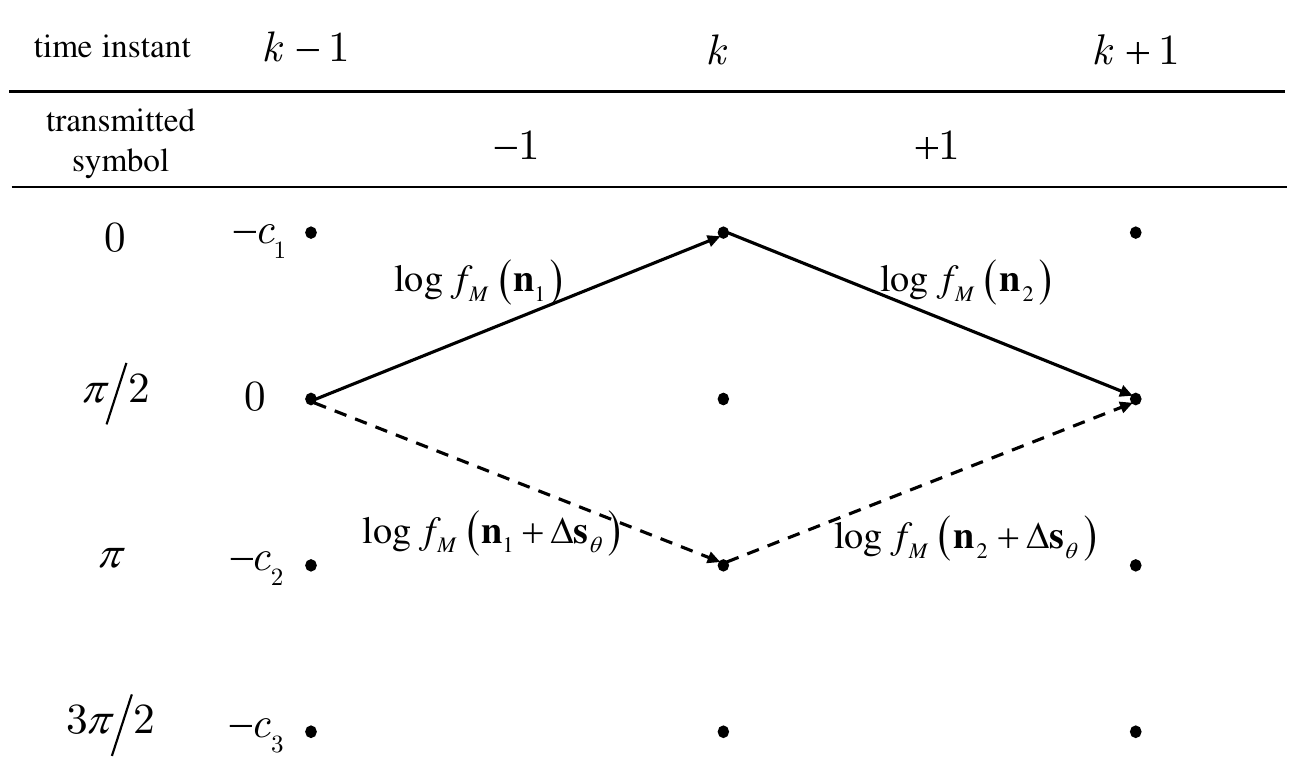}
\caption{The error event of MSK demodulation based on VA with the largest probability}
\label{MSK_event_explain}
\end{figure}

For MSK modulation, the VA can be described by 4-state trellis. As demonstrated in \cite{paper21}, the error event shown in Fig. \ref{MSK_event_explain} has the largest probability.

In Fig. \ref{MSK_event_explain}, we assume that the transmitted symbols for time instant $k$ and $k+1$ are $(-1,+1)$ and the corresponding branch metrics are presented. The correct path is presented by solid line. The estimated path is denoted by dashed line, which represents symbols $(+1,-1)$ for time instant $k$ and $k+1$. The resulting probability of the error event is used to approximate the theoretical BER. It is given as
\begin{align}\label{joint_probability_MSK}
&P_{e,MSK}\approx P_{J,p}\nonumber\\
=&P\big(\log f_M(\mathbf{n}_1+\Delta\mathbf{s}_{\theta})>\log f_M(\mathbf{n}_1),\log f_M(\mathbf{n}_1+\Delta\mathbf{s}_{\theta})\nonumber\\
&+\log f_M(\mathbf{n}_2+\Delta\mathbf{s}_{\theta})>\log f_M(\mathbf{n}_1)+\log f_M(\mathbf{n}_2)\big)\nonumber\\
=&P\big(\log f_M(\mathbf{n}_1+\Delta\mathbf{s}_{\theta})+\log f_M(\mathbf{n}_2+\Delta\mathbf{s}_{\theta})>\log f_M(\mathbf{n}_1)\nonumber\\ &+\log f_M(\mathbf{n}_2)|f_M(\mathbf{n}_1+\Delta\mathbf{s}_{\theta})>f_M(\mathbf{n}_1)\big)\nonumber\\
&\times P\big(\log f_M(\mathbf{n}_1+\Delta\mathbf{s}_{\theta})>\log f_M(\mathbf{n}_1)\big)\nonumber\\
\leq& P\big(\log f_M(\mathbf{n}_1+\Delta\mathbf{s}_{\theta})>\log f_M(\mathbf{n}_1)\big)\triangleq P^U_{e,MSK},
\end{align}
where $\Delta\mathbf{s}_{\theta}=\mathbf{s}_{\theta}(d=+1)-\mathbf{s}_{\theta}(d=-1)$ and $d$ denotes the transmitting symbol. For example, if the starting state is phase $0$ at time instant $k-1$, the phase of ending state at time instant $k$ can be $\pi/2$ or $3\pi/2$, which corresponds to two paths. The corresponding MSK signals are denoted as $\mathbf{s}_{\theta}(d=+1)$ and $\mathbf{s}_{\theta}(d=-1)$. Note that the subscript $p$ of $P_{J,p}$ denotes the dimension of the considered noise PDF. Since the conditional probability $P_{J,p}$ is less than 1 and then, we set $P^U_{e,MSK}$ as the upper bound of the theoretical BER.

For the lower bound, we need to calculate the $P_{J,p}$ by averting the difficulty due to the high-dimensional integral. \eqref{joint_probability_MSK} can be further expressed as
\begin{align}\label{joint_expression_high_dimension}
P_{J,p}=&\int_{\Omega_{M}}f_M(\mathbf{n}_1)\int_{\tilde{\Omega}_{M}(\mathbf{n}_1)}f_M(\mathbf{n}_2)d\mathbf{n}_2d\mathbf{n}_1,
\end{align}
where $\Omega_{M}=\{\mathbf{n}_1:f_M(\mathbf{n}_1+\Delta\mathbf{s}_{\theta})>f_M(\mathbf{n}_1)\}$ and $\tilde{\Omega}_{M}(\mathbf{n}_1)=\{\mathbf{n}_2:\log f_M(\mathbf{n}_2+\Delta\mathbf{s}_{\theta})+\log f_M(\mathbf{n}_1+\Delta\mathbf{s}_{\theta})-\log f_M(\mathbf{n}_1)>\log f_M(\mathbf{n}_2)\}$. For convenience, we denote
\begin{align}\label{Delta_f_M}
\Delta f_M(\mathbf{n}_1)= \log\left[\frac{f_M(\mathbf{n}_1+\Delta\mathbf{s}_{\theta})}{f_M(\mathbf{n}_1)}\right],
\end{align}
\begin{align}\label{Omega_M_A}
\tilde{\Omega}_{M}^A(\mathbf{n}_1)=\{\mathbf{n}_2:\Delta f_M(\mathbf{n}_2)+\Delta f_M(\mathbf{n}_1)>0,\rho=1\},
\end{align}
\begin{align}\label{Omega_M_B}
\tilde{\Omega}_{M}^B(\mathbf{n}_1)=\{\mathbf{n}_2:\Delta f_M(\mathbf{n}_2)+\Delta f_M(\mathbf{n}_1)>0,\rho=0\}.
\end{align}

\begin{figure}[htbp]
\centering
\includegraphics[width=3.4in]{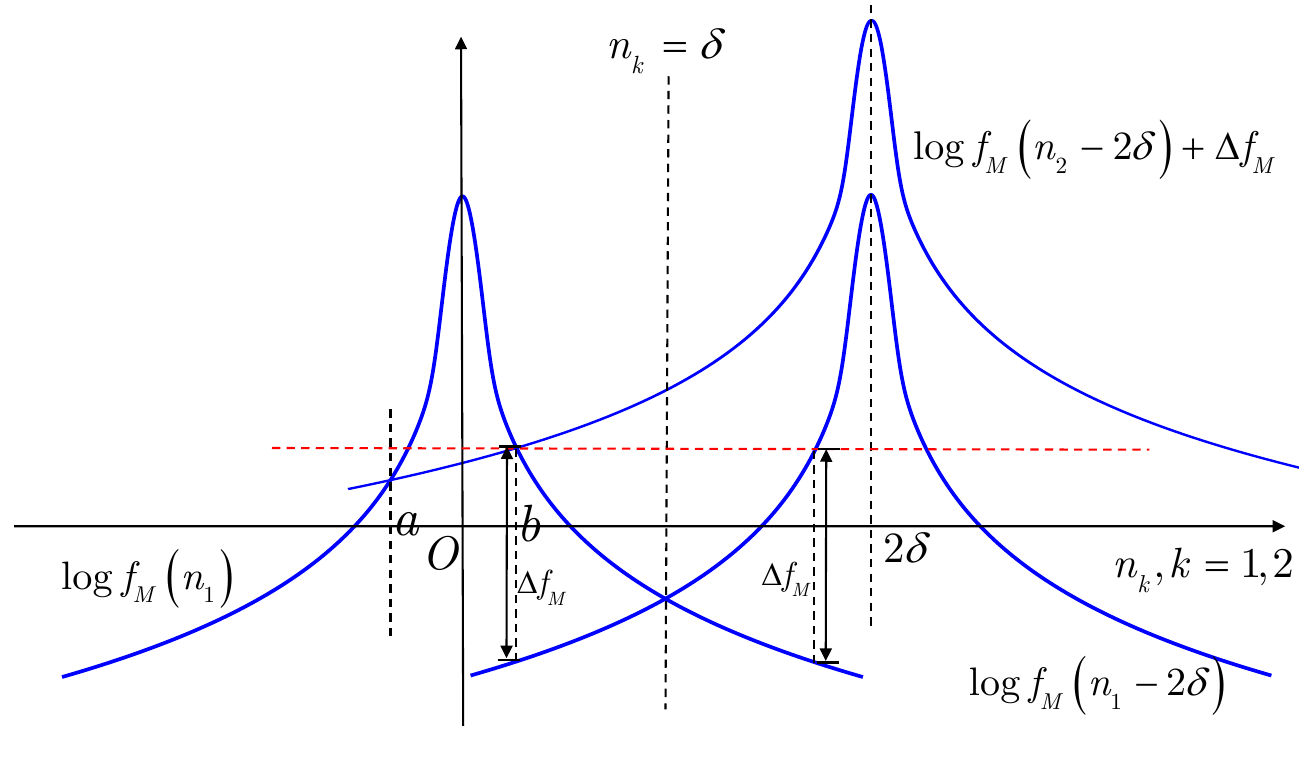}
\caption{The one-dimensional sketch of the conditional probability $P_{J,p}$}
\label{MSK_BER_explain}
\end{figure}

For $f_M(\mathbf{n}_1)$ in \eqref{joint_expression_high_dimension}, the exponential term of $f_M(\mathbf{n}_1)$ can be omitted when the integral region $\Omega_{M}$ is far from the origin or equivalently, the signal power is large. However, $\tilde{\Omega}_{M}(\mathbf{n}_1)$ can be very close to or even contains the origin since $\Delta f_M\geq0$. Therefore, omitting the exponential term of $f_M(\mathbf{n}_2)$ will introduce large error to $P_{J,p}$. As it is very difficult to determine $\tilde{\Omega}_{M}(\mathbf{n}_1)$ analytically, we first describe how to approximate the joint probability in \eqref{joint_probability_MSK}. The key idea is illustrated in Fig. \ref{MSK_BER_explain} by taking one dimension as an example.

In Fig. \ref{MSK_BER_explain}, we denote the one-dimensional version of $\Delta\mathbf{s}_{\theta}$ by $-2\delta$. The joint probability is equivalent to
\begin{align}
P_{J,1}=&P\big(\log f_M(n_1-2\delta)>\log f_M(n_1),\log f_M(n_2-2\delta)\nonumber\\
&+\underbrace{\log f_M(n_1-2\delta)-\log f_M(n_1)}_{\Delta f_M(n_1)>0}>\log f_M(n_2)\big),
\end{align}
where $\Delta f_M(n_1)>0$ always holds since $n_1\in\Omega_M=(\delta,+\infty)$. If $n_1\in(\delta,2\delta]$ and $\Delta f_M(n_2)+\Delta f_M(n_1)>0$ holds, the range of $n_2$ should be $(-\infty,a)\cup(b,+\infty)$ and $b=2\delta-n_1$. It can be observed that $a=b=0$ when $n_1=2\delta$. Although $a$ could not be determined in general, it can be found that $a$ will be close to $b$ as $\delta$ increases. Specifically, $\lim\limits_{\delta\rightarrow+\infty}a=b$. The reason is that $\log f_M(n_1)\sim\mathcal{O}(-\log n_1)$ and its slope converges to 0 if $n_1\rightarrow+\infty$. Therefore, we adopt the approximation $a=b$. Note that this approximation is more accurate for larger SNR because $\delta$ is related to the signal power. Meanwhile, similar approximation can be applied for the case of $n_1\in[2\delta,+\infty]$. Based on the above discussion, we can have
\begin{align}\label{joint_expression_one_dimension}
P_{J,1}= &\int_{\delta}^{+\infty}f_M(n_1)\int_{|2\delta-n_1|}^{+\infty}f_M(n_2)dn_2dn_1.
\end{align}

\begin{figure}[htbp]
\centering
\includegraphics[width=3.4in]{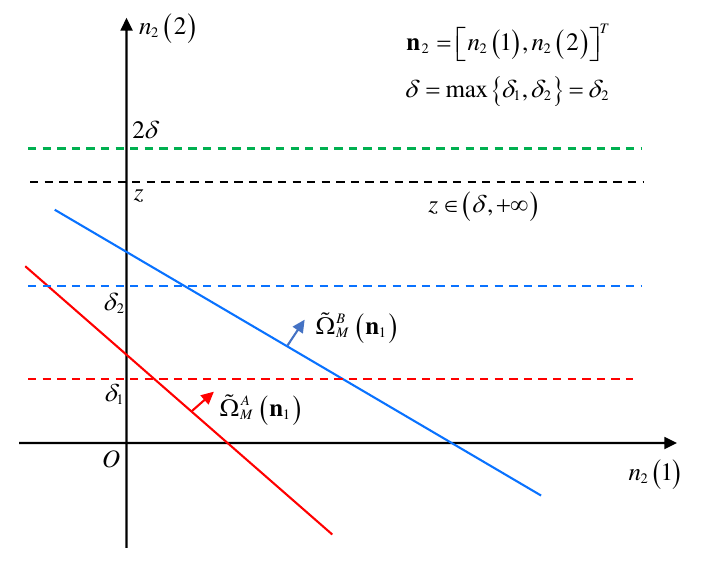}
\caption{The high-dimensional sketch of the conditional probability $P_{J,p}$ with $p=2$}
\label{MSK_BER_explain_2}
\end{figure}

For the case of $p>1$, the integral region can be rotated such that the corresponding hyperplane is perpendicular to some coordinate axis. As illustrated in the Fig. \ref{MSK_BER_explain_2}, let the intersection point be $\delta_1$ for $\tilde{\Omega}_{M}^A(\mathbf{n}_1)$ and $\delta_2$ for $\tilde{\Omega}_{M}^B(\mathbf{n}_1)$ after rotation, respectively. Note that $P_{J,p}$ will become smaller when SNR increases, i.e., $\delta$ becomes larger. We set $\delta=\max\{\delta_1,\delta_2\}$ to obtain the lower bound of $P_{J,p}$. The dashed line represents the hyperplane of integral regime corresponding to the rotation version of original integral region. Indeed, the vertical axis in Fig. \ref{MSK_BER_explain_2} can be considered as the horizontal axis in Fig. \ref{MSK_BER_explain}. Then, we can extend the one-dimensional expression to higher dimension case with the aid of Fig. \ref{MSK_BER_explain} and \eqref{joint_expression_one_dimension}. With a slight abuse of notation, we replace $n_2(1)$ and $n_2(2)$ by $n_1$ and $n_2$ for convenience. Furthermore, it follows Appendix \ref{appendix_theorem_3} that  \eqref{joint_expression_high_dimension} could be converted to
\begin{align}\label{joint_expression_high_dimension_2}
P_{J,p}\geq&\int_{\delta}^{2\delta}T(n_1)\int_{|2\delta-n_1|}^{+\infty}\left[\frac{\rho}{2\gamma_g\sqrt{\pi}}\exp\left(-\frac{n_2^2}{4\gamma_g^2}\right)+T(n_2)\right]\nonumber\\
&\times dn_2dn_1\triangleq P_{J,p}^L,
\end{align}
where
\begin{align}
T(x)\triangleq\int_{0}^{+\infty}\frac{(1-\rho)k_2\Upsilon(p)\sqrt{\det\left(\Sigma\right)}r^{p-2}}{\left(1+\left(r^2+x^2\right)/\alpha\right)^\frac{\alpha+p}{2}}dr.
\end{align}

Then, we have the following theorem for the BER bounds of MSK signals.
\begin{theorem}\label{theorem_5}
Let the received signal model be \eqref{MSK_received_signal_model} and $K=p$. The theoretical BER $P_{e,MSK}$ of MSK signals demodulated via VA is bounded as follows,
\begin{align}\label{MSK_upper_bound_nonfading}
P_{e,MSK}\leq\frac{1}{4}\sum_{\theta\in\Omega_{\theta}}P_{e,MSK}^U(\Delta\mathbf{s}_{\theta}),
\end{align}
\begin{align}\label{MSK_lower_bound_nonfading}
P_{e,MSK}\geq\frac{1}{4}\sum_{\theta\in\Omega_{\theta}}P_{e,MSK}^L(\Delta\mathbf{s}_{\theta}),
\end{align}
where the expressions of $P_{e,MSK}^U(\Delta\mathbf{s}_{\theta})$ and $P_{e,MSK}^L(\Delta\mathbf{s}_{\theta})$ are given as \eqref{upper_bound_MSK_given_s_i} and \eqref{lower_bound_given_delta_s_i}, respectively.
\end{theorem}
\begin{IEEEproof}
The proof is relegated to Appendix \ref{appendix_theorem_5}.
\end{IEEEproof}

$\mathbf{Remarks}$: In fact, if we denote the noise vectors of two adjacent MSK symbol periods by $\mathbf{n}_1$ and $\mathbf{n}_2$, it can be observed that $\mathbf{n}_1$ and $\mathbf{n}_2$ are not completely mutually independent. In other words, $f_M(\mathbf{n}_2)$ is not symmetric about the origin and the analysis based on Fig. \ref{MSK_BER_explain} is biased. However, the symbol period length equals to the noise memory order $p$ and the time intervals of most samples in $\mathbf{n}_1$ and $\mathbf{n}_2$ are larger than $p$. Therefore, we can omit the correlation between noise samples in different symbol periods. Moreover, the approximation error brought by $a\approx b$ in Fig. \ref{MSK_BER_explain} will vanish as the signal power increases, which implies the asymptotic behavior of the proposed bounds.

\section{simulations}\label{simulation}
In this section, we evaluate the demodulation performance and verify the theoretical BER analysis under bursty mixed noise. The generalized signal-to-noise ratio (GSNR) is adopted since the bursty minted noise has no finite second moment.The GSNR is defined as
\begin{equation}\label{GSNR}
\text{GSNR(dB)}=10\log_{10}\frac{P_s}{2(\gamma_g^2+\gamma_s^2)},
\end{equation}
where the $P_s$ is the power of the transmitting signals.

To investigate the influence of the parameters $\alpha$, $\rho$ and $p$ on the demodulation performance, we fix $\gamma_s=\gamma_g=2$. Meanwhile, $\alpha$ is set to be 1.2 and 1.8 to represent the noise with large impulsiveness and close to WGN, respectively. The regularized covariance matrix is set to be $\tilde{\Sigma}(1,:)=[1,0.7]$ when $p=2$ and $\tilde{\Sigma}(1,:)=[1,0.8,0.6,0.4,0.2]$ when $p=5$. Note that the other elements of $\tilde{\Sigma}$ are correspondingly determined as it is a Topelize matrix.

\subsection{Demodulation performance comparison}
We first provide some conventional demodulation metrics used for performance comparison. Here, we present the baselines for BPSK. The expressions associated with the other modulation schemes can be similarly obtained by following section \ref{application} and \ref{section_demodulation_MSK}.
\begin{enumerate}
\item{$L_1$ norm demodulator ($L_1$D): The metric is $L_1$ norm and the demodulation problem can be expressed as
\begin{equation}
\hat{x}_i=\mathop{\arg\max}\limits_{x_i\in\Omega(2)}-\sum_{j=1}^{K}|y_i(j)-x_i(j)|.
\end{equation}
}
\item{$L_2$ norm demodulator ($L_2$D): The metric is Euclidean distance and the demodulation problem can be expressed as
\begin{equation}
\hat{x}_i=\mathop{\arg\max}\limits_{x_i\in\Omega(2)}-\sum_{j=1}^{K}[y_i(j)-x_i(j)]^2.
\end{equation}}
\item{Myriad demodulator (MyD) \cite{paper18}: Myriad detector is optimal with respect to white Cauchy noise. Moreover, it is also an adaptive and suboptimal metric for S$\alpha$S noise. The signal demodulation can be expressed as
\begin{equation}
\hat{x}_i=\mathop{\arg\max}\limits_{x_i\in\Omega(2)}-\sum_{j=1}^{K}\log\left[\kappa^2+\frac{(y_i(j)-x_i(j))^2}{\gamma_s^2}\right],
\end{equation}
    where the $\kappa=\sqrt{\alpha/(2-\alpha)}$.}
\item{$\alpha$SG demodulator ($\alpha$SGD) \cite{paper17}: $\alpha$SG demodulator is proposed with respect to the bursty IN based on the $\alpha$SG model. However, due to the lack of analytical PDF, specific approximation has to be utilized to avoid prohibitively high complexity of integral. For example,
\begin{align}
\hat{x}_i=&\mathop{\arg\max}\limits_{x_i\in\Omega(2)}-\sum_{j=1}^{p}\log\left[y_i(j)-x_i(j)-\mu_{j1}\right]\nonumber\\
&-\sum_{j=p+1}^{K}\log\left[y_i(j)-x_i(j)-\mu_{j2}\right],
\end{align}
    where the expression of $\mu_{j1}$ and $\mu_{j2}$ can be found in \cite{paper17}.}
\item{ML demodulator without memory (MLD(W)): By ignoring the correlation between the noise samples, this method directly utilizes the PDF in \eqref{one_dimension} to perform the demodulation, i.e.,
\begin{equation}
\hat{x}_i=\mathop{\arg\max}\limits_{x_i\in\Omega(2)}\sum_{j=1}^{K}f_M(y_i(j)-x_i(j),p=1).
\end{equation}}
\end{enumerate}

In the simulations, we set the oversampling rate to be $K=20$ and focus on the influence of $\alpha$ and $\rho$ on the demodulation performance.
\begin{figure}[htbp]
\centering
\subfloat[$\alpha=1.2$, $\rho=0.2$]{\includegraphics[width=1.75in]{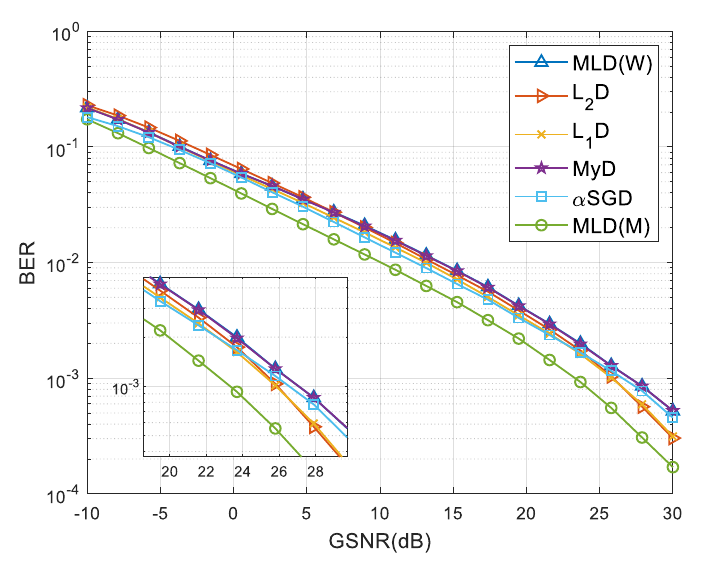}\label{fig_4_a}}
\subfloat[$\alpha=1.2$, $\rho=0.8$]{\includegraphics[width=1.75in]{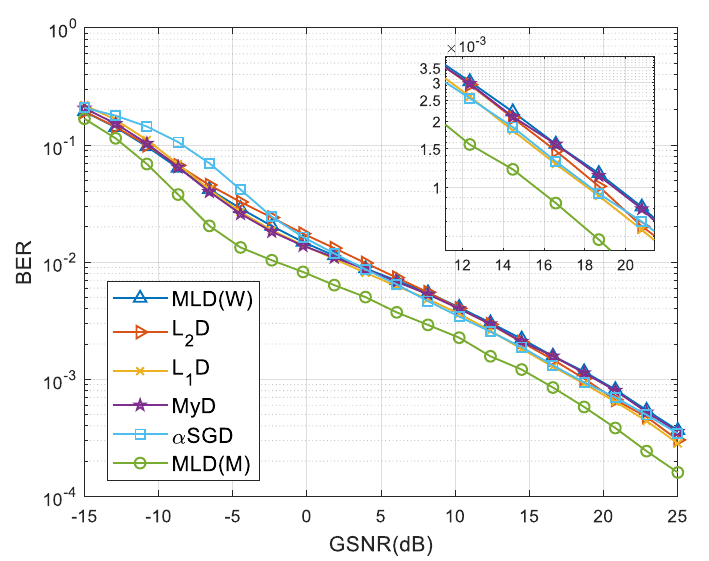}\label{fig_4_b}}

\subfloat[$\alpha=1.8$, $\rho=0.2$]{\includegraphics[width=1.75in]{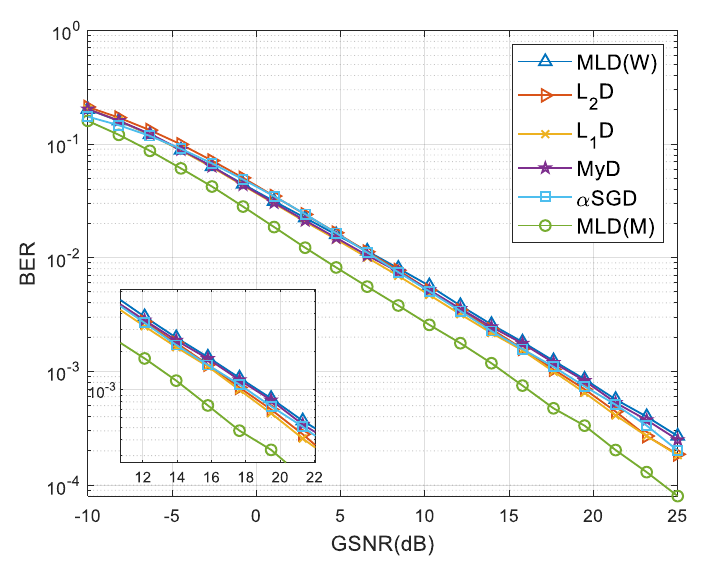}\label{fig_4_c}}
\subfloat[$\alpha=1.8$, $\rho=0.8$]{\includegraphics[width=1.75in]{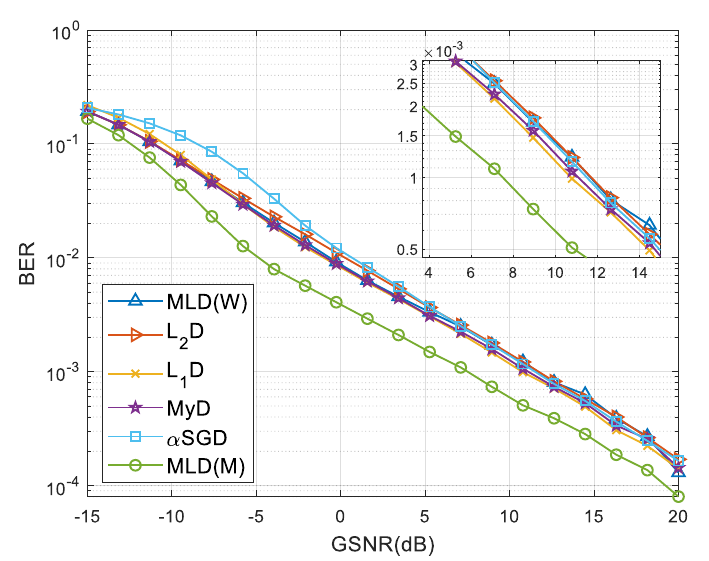}\label{fig_4_d}}
\caption{The BPSK demodulation performance comparison under bursty mixed noise}
\label{fig_4}
\end{figure}

In Fig. \ref{fig_4}, we present the demodulation performance comparison for different demodulation methods under various bursty mixed noise scenarios. We denote the ML demodulation considering memory in \eqref{likelihood_function} by `MLD(M)'. The simulation results showcase that `MLD(M)' outperforms other methods with a performance gain of more than 2.5 dB for BER $10^{-3}$ in all scenarios. The norm, myriad and MLD(W) based methods have similar BER performance since the correlation between noise samples is omitted. Even though the `$\alpha$SGD' based method takes into account the thick tail feature of noise PDF and the correlation between noise samples, it ignores the background Gaussian noise, which leads to model mismatch and performance deterioration especially when the GSNR is relatively low. Meanwhile, from Fig.\ref{fig_4_b} and \ref{fig_4_d}, it can be observed that the BER performance decreases first quadratically and then linearly. The underlying reason is that the BER is mainly influenced by WGN in the low GSNR regions. When the GSNR increases, the demodulation performance is mainly affected by the IN cluster.

\begin{figure}[htbp]
\centering
\subfloat[$\alpha=1.2$, $\rho=0.2$]{\includegraphics[width=1.75in]{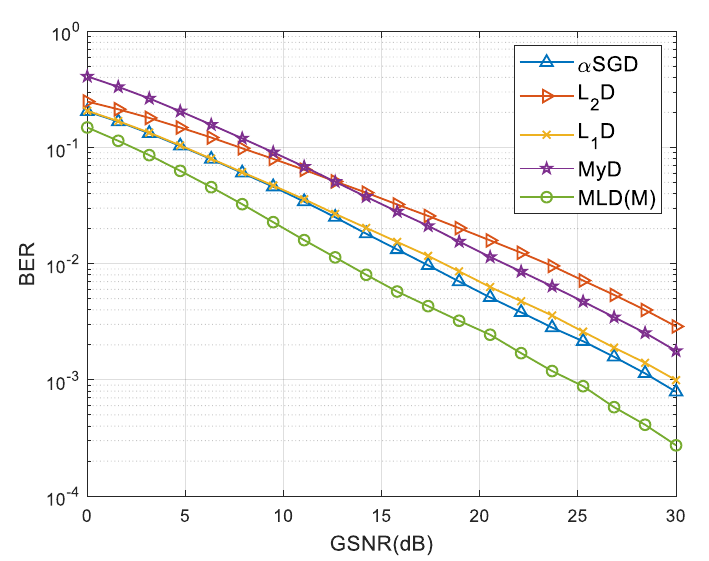}\label{fig_9_a}}
\subfloat[$\alpha=1.8$, $\rho=0.8$]{\includegraphics[width=1.75in]{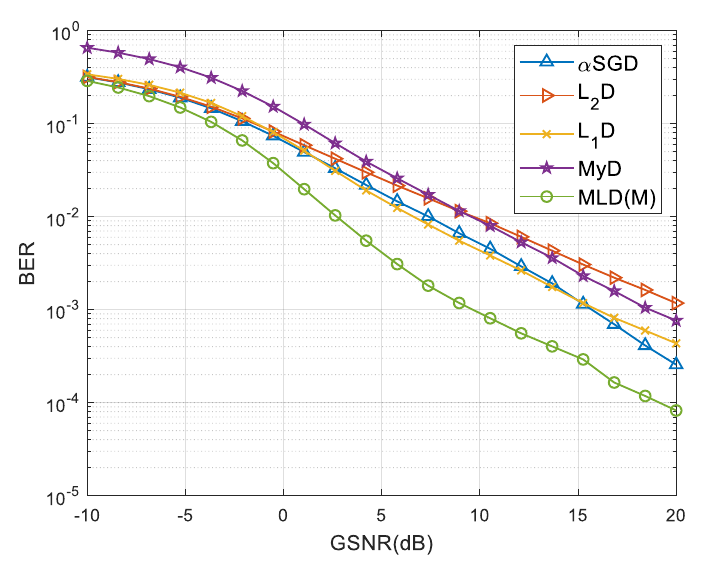}\label{fig_9_b}}
\caption{The 8-PSK demodulation performance comparison between the proposed algorithm and conventional methods in the bursty mixed noise}
\label{fig_9}
\end{figure}

Subsequently, the BER of 8-PSK is simulated to validate the proposed algorithm based on the complex GS model. Note that $\alpha$ and $\rho$ have similar influence on the performance, and the impulsive feature of noise will become larger when $\alpha$ and $\rho$ are smaller. We consider $\alpha=1.2,\rho=0.2$ and $\alpha=1.8,\rho=0.8$, which separately represent the scenarios with higher and lower impulsive feature, respectively. As the performance of `MLD(W)' is akin to the other baselines, we omit it in what follows. From Fig. \ref{fig_9}, it can be seen that the performance gain of our proposed `MLD(M)' over the baselines is more considerable compared with BPSK modulation scheme. The gain is about 3.8dB when BER reaches $10^{-3}$. The `$L_1$D' has comparable performance with `$\alpha$SGD', which has the best BER performance among conventional methods. The Euclidean distance based metric is not effective especially when GSNR is large. For the scenario with large GSNR, the main component of the mixed noise is the bursty IN. Note that the detrimental influence of IN is more considerable than that of WGN if they have the same power. This is because the power of IN is determined by the successive outliers. Accordingly, `MyD' performs better than `$L_2$D' in large GSNR region and worse in small GSNR region.

\begin{figure}[htbp]
\centering
\subfloat[$\alpha=1.2$, $\rho=0.2$]{\includegraphics[width=1.75in]{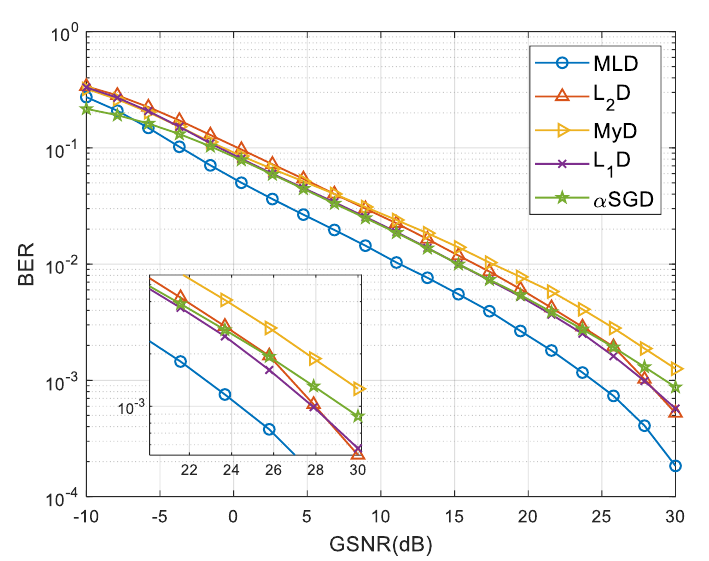}\label{fig_10_a}}
\subfloat[$\alpha=1.8$, $\rho=0.8$]{\includegraphics[width=1.75in]{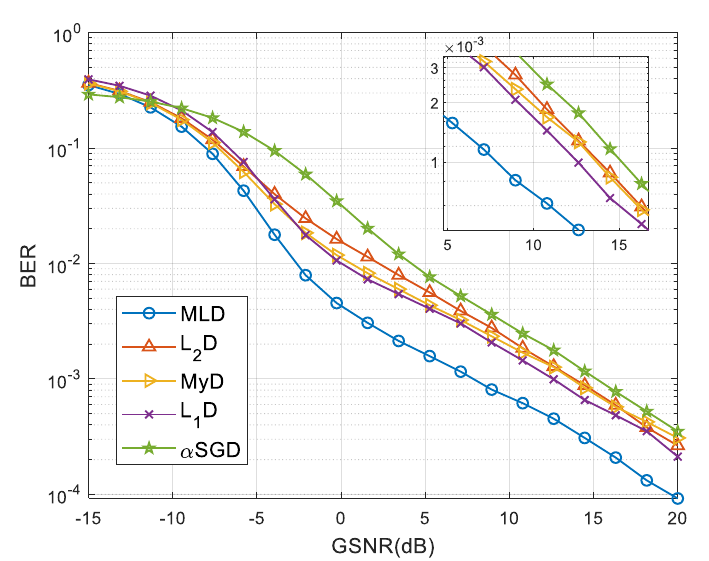}\label{fig_10_b}}
\caption{The MSK demodulation performance comparison between the proposed algorithm and conventional methods in the bursty mixed noise}
\label{fig_10}
\end{figure}

In Fig. \ref{fig_10}, the demodulation performance of MSK signals is compared. It can be observed that `MLD(M)' outperforms the other methods with gain about 4dB when BER is $10^{-3}$. Meanwhile, the `$\alpha$SGD' has the best performance among the baselines when the impulsive feature is significant since $\alpha$SG model is proposed for the pure bursty IN. When $\alpha$ and $\rho$ increase, the performance of `$\alpha$SGD' deteriorates since it ignores the background WGN.

\subsection{Verification of Theoretical BER analysis}
\begin{figure}[htbp]
\centering
\subfloat[$p=1,\rho=0.2$]{\includegraphics[width=1.75in]{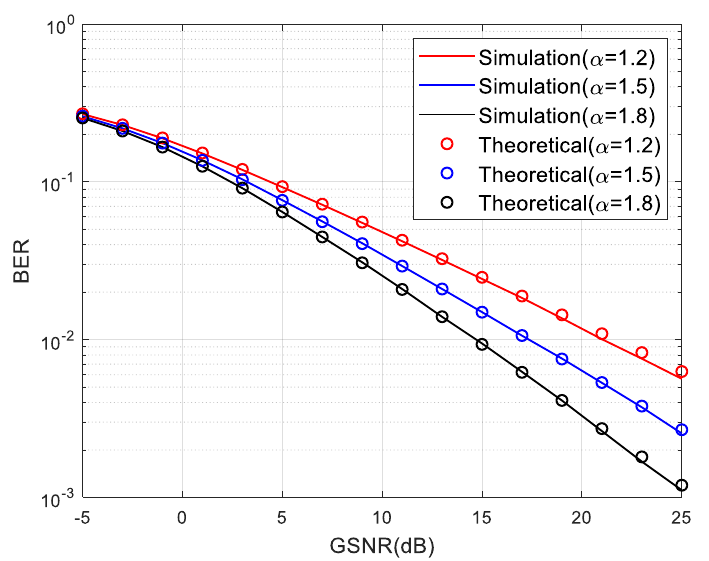}\label{fig_5_a}}
\subfloat[$p=1,\rho=0.8$]{\includegraphics[width=1.75in]{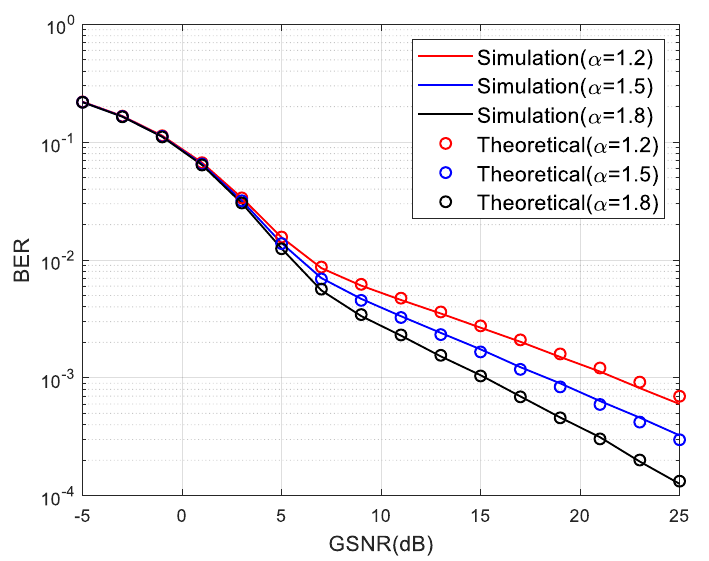}\label{fig_5_b}}
\caption{The theoretical BER and the simulated BER comparison of BPSK scheme in the bursty mixed noise with $p=1$ and various $\alpha,\rho$}
\label{fig_5}
\end{figure}

The theoretical BER analysis is first verified for $p=1$, for which the noise becomes white mixed noise. The other simulation setup is the same as that for Fig. \ref{fig_4}. The BER of BPSK is presented in Fig. \ref{fig_5}. It can be seen that the theoretical BER matches the simulated BER in the whole GSNR region.

\begin{figure}[htbp]
\centering
\subfloat[$p=5,\rho=0.2$]{\includegraphics[width=1.75in]{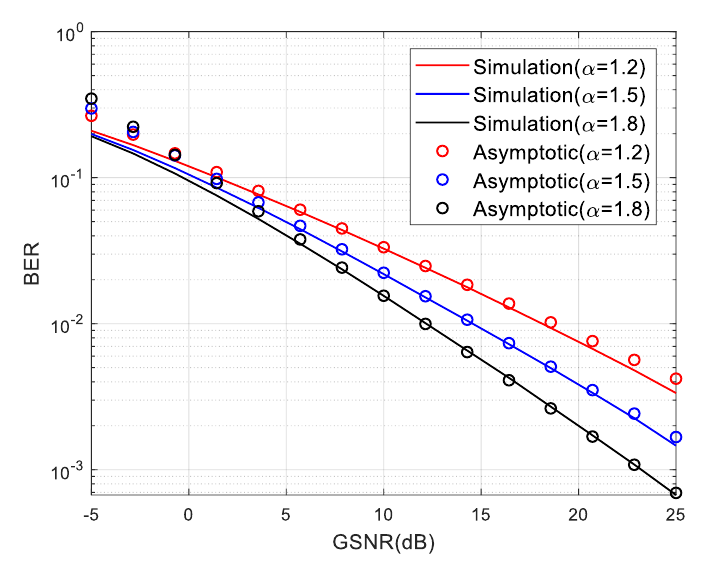}\label{fig_5_e}}
\subfloat[$p=5,\rho=0.8$]{\includegraphics[width=1.75in]{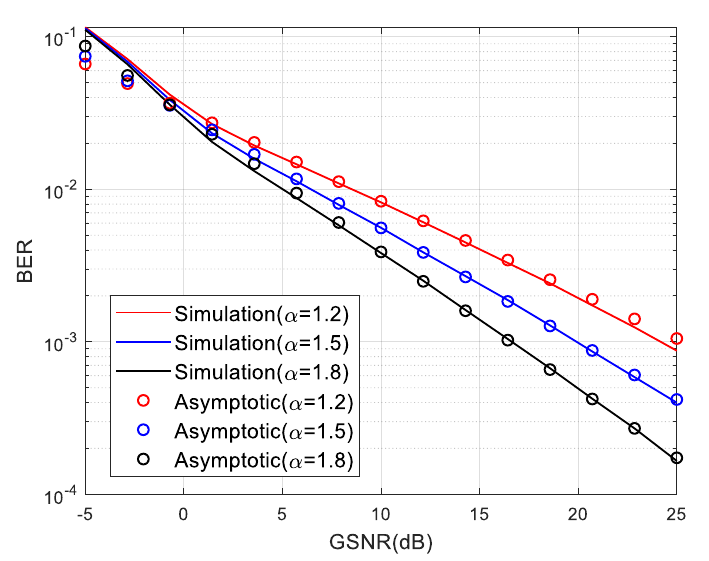}\label{fig_5_f}}
\caption{The theoretical asymptotic BER and the simulated BER comparison of BPSK scheme in the bursty mixed noise with $p=5$ and various $\alpha,\rho$}
\label{fig_6}
\end{figure}

Furthermore, the demodulation performance of BPSK and its asymptotic BER $\hat{P}_e$ with $p\geq2$ are presented in Fig. \ref{fig_6}. It can be observed that the asymptotic BER is very tight for relatively large GSNR. In fact, $\hat{P}_e$ will converges to real error probability with the increase of GSNR.

\begin{figure*}[htbp]
\centering
\subfloat[$\alpha=1.2,\rho=0.2$]{\includegraphics[width=1.75in]{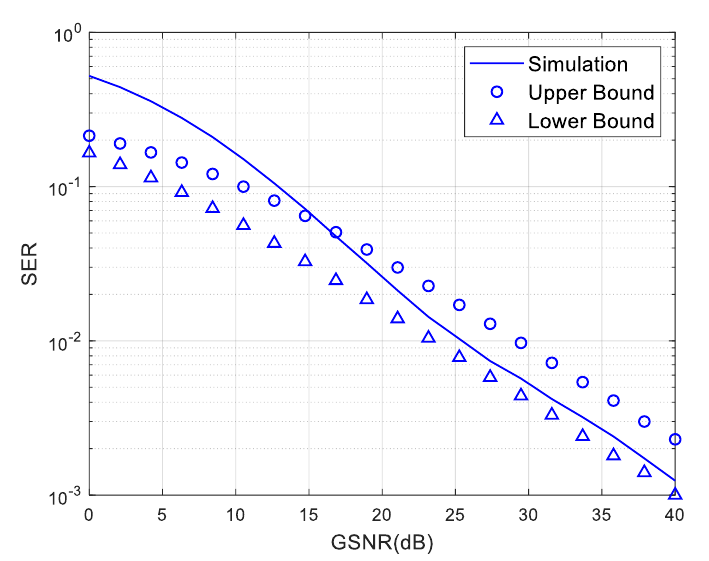}\label{fig_7_a}}
\subfloat[$\alpha=1.2,\rho=0.8$]{\includegraphics[width=1.75in]{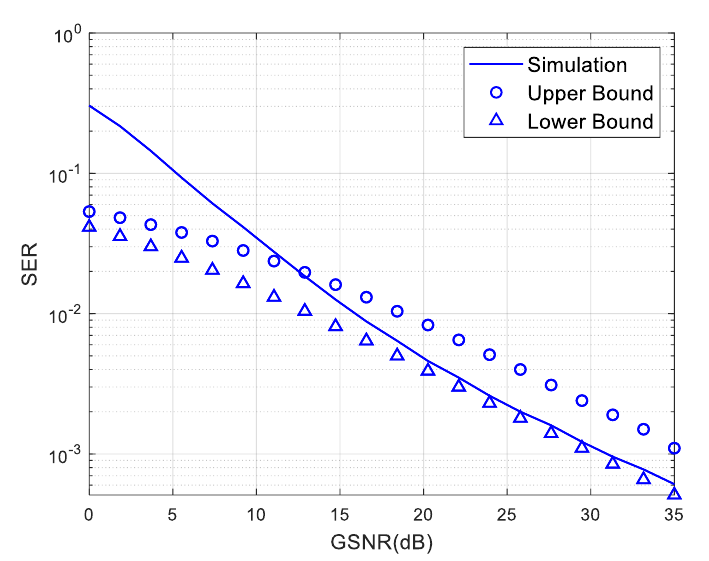}\label{fig_7_b}}
\subfloat[$\alpha=1.8,\rho=0.2$]{\includegraphics[width=1.75in]{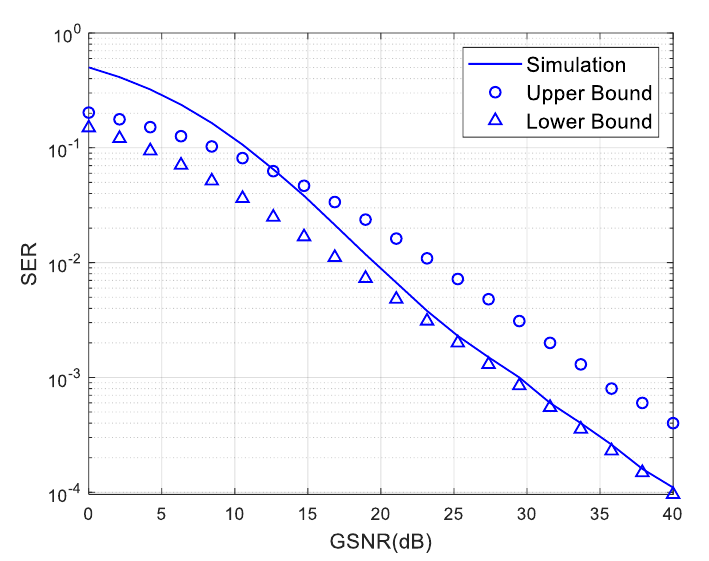}\label{fig_7_c}}
\subfloat[$\alpha=1.8,\rho=0.8$]{\includegraphics[width=1.75in]{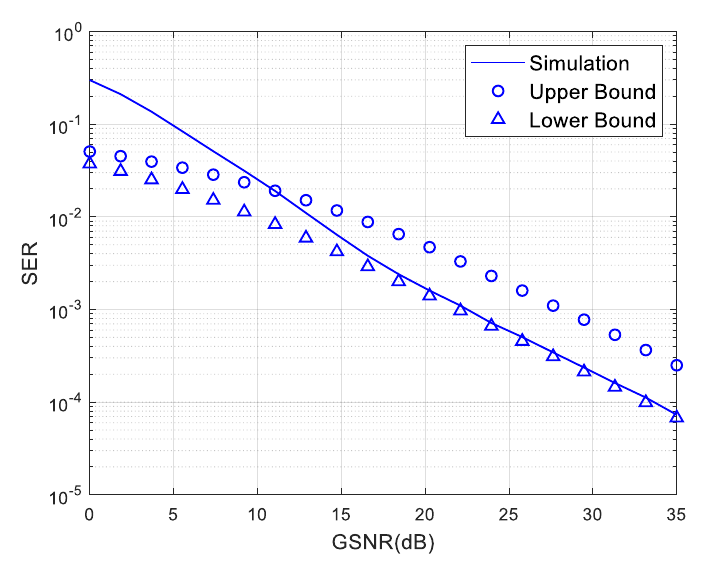}\label{fig_7_d}}
\caption{Comparison between the theoretical SER bounds and the simulated SER for 8-PSK modulation scheme in the bursty mixed noise with $p=5$ and various $\alpha,\rho$.}
\label{fig_7}
\end{figure*}

The SER bounds of 8-PSK are compared with the simulated SER in Fig. \ref{fig_7}. It can be seen that the lower bound is rather tight for various $\alpha$. The upper bound is more efficient when noise impulsiveness is stronger. This phenomenon is consistent with \eqref{expression_A_ikl}.

\begin{figure*}[htbp]
\centering
\subfloat[$\alpha=1.2,\rho=0.2$]{\includegraphics[width=1.75in]{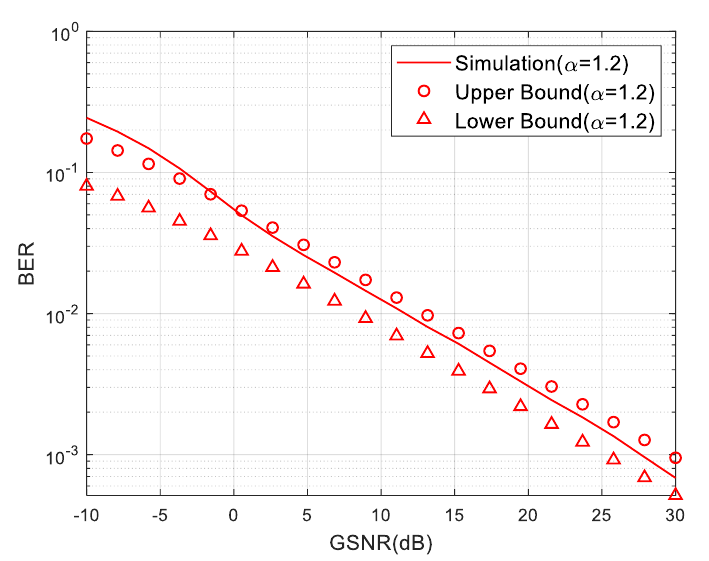}\label{fig_8_a}}
\subfloat[$\alpha=1.2,\rho=0.8$]{\includegraphics[width=1.75in]{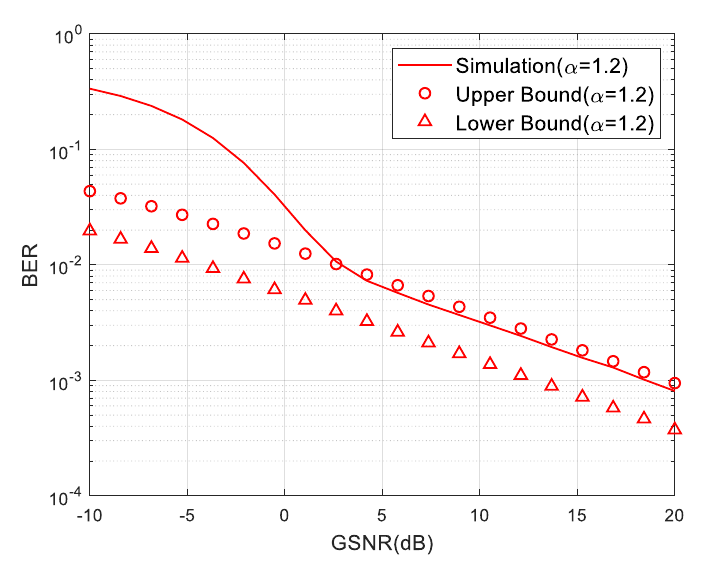}\label{fig_8_b}}
\subfloat[$\alpha=1.8,\rho=0.2$]{\includegraphics[width=1.75in]{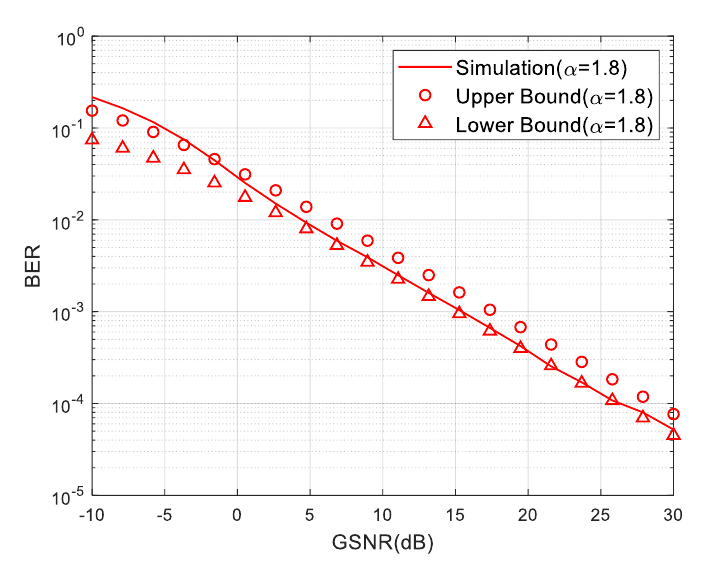}\label{fig_8_a}}
\subfloat[$\alpha=1.8,\rho=0.8$]{\includegraphics[width=1.75in]{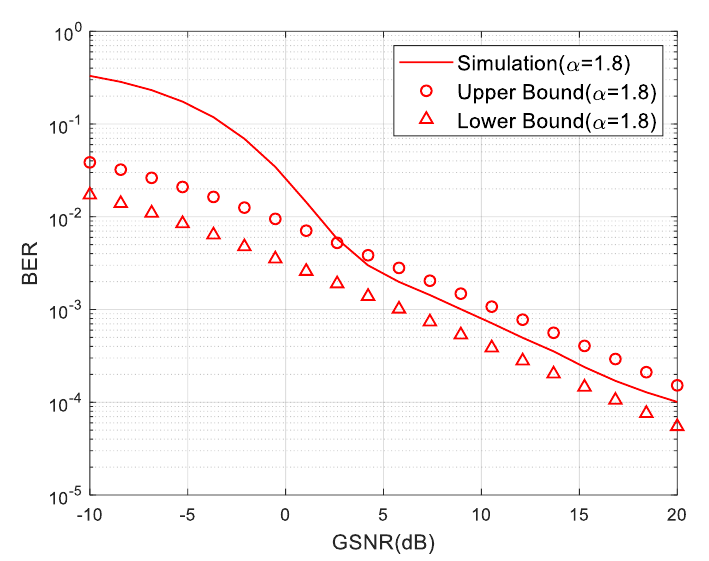}\label{fig_8_b}}
\caption{The comparison of bounds and the simulated BER of MSK scheme in the bursty mixed noise with $p=5$ and various $\alpha,\rho$.}
\label{fig_8}
\end{figure*}

Finally, the proposed BER bounds of MSK are compared with the simulated BER in Fig. \ref{fig_8}. It can be observed that the bounds are rather tight in different noise scenarios. This result verifies the difference between the real BER and the probability of error event presented in Fig. \ref{MSK_event_explain} is small. Note that the bounds are asymptotic due to the simplification of PDF. Consequently, the performance bounds are tight in the high GSNR region. These results validate the theoretical analysis.

\section{conclusion}\label{conclusion}
In the second part, based on the PDF proposed in the first part of the paper, the M-PSK and MSK demodulation algorithms based on the ML criteria were designed. The theoretical BER performance of M-PSK scheme was discussed for different cases. As for the MSK scheme, we derived a upper and lower bound for the demodulation BER based on VA. Numerical results demonstrated that the GS model based ML demodulation has significant performance gain compared with the baselines. The theoretical BER matched the simulation BER well, and the error probability bounds were also tight under various scenarios.

\begin{appendices}

\section{Proof of theorem \ref{theorem_3}}\label{appendix_theorem_3}
\setcounter{equation}{0}
\renewcommand{\theequation}{A.\arabic{equation}}
We give the detailed derivation of the asymptotic BER expression based on the coordinate transformation. The theoretical BER is
\begin{align}
\hat{P}_e=\int_{\Omega_B}\frac{(1-\rho)k_2}{\left(1+\Vert\Sigma^{-1/2}\mathbf{n}\Vert^2/\alpha\right)^\frac{\alpha+p}{2}}d\mathbf{n},
\end{align}
where we denote $\Omega_B=\{\mathbf{n}:(\Sigma^{-1}\mathbf{e}_p)^\top\mathbf{n}>P_0\}$ and $P_0=A\mathbf{e}_p^\top\Sigma^{-1}\mathbf{e}_p$. Let $\mathbf{x}=\Sigma^{-1/2}\mathbf{n}$, we have
\begin{align}
\hat{P}_e=\det\left(\Sigma^{-1/2}_{inv}\right)\int_{\Omega_C}\frac{(1-\rho)k_2}{\left(1+\Vert\mathbf{x}\Vert^2/\alpha\right)^\frac{\alpha+p}{2}}d\mathbf{x},
\end{align}
where $\Sigma^{-1/2}_{inv}$ represents the inverse of $\Sigma^{-1/2}$ and $\Omega_C=\{\mathbf{x}:(\Sigma^{-1}\mathbf{e}_p)^\top\Sigma^{-1/2}_{inv}\mathbf{x}>P_0\}$.

It is difficult to calculate the integral directly with respect to the region $\Omega_C$. To avert this difficulty, we utilize the Given's transform to assure that the rotated hyperplane $\{\mathbf{x}:(\Sigma^{-1}\mathbf{e}_p)^\top\Sigma^{-1/2}_{inv}\mathbf{Q}\mathbf{x}=P_0\}$ is paralleled to some axis. Without loss of generality, we choose the rotation matrix $\mathbf{Q}\in\mathbb{R}^{p\times p}$ to satisfy
\begin{align}\label{rotation_matrix}
(\Sigma^{-1}\mathbf{e}_p)^\top\Sigma^{-1/2}_{inv}\mathbf{Q}=[\lambda,0,\cdots,0],
\end{align}
and set $\mathbf{y}=\mathbf{Q}^{-1}\mathbf{x}$. Thus,
\begin{align}\label{P_e_Omega_D}
\hat{P}_e=\det\left(\Sigma^{-1/2}_{inv}\mathbf{Q}\right)\int_{\Omega_D}\frac{(1-\rho)k_2}{\left(1+\Vert\mathbf{y}\Vert^2/\alpha\right)^\frac{\alpha+p}{2}}d\mathbf{y},
\end{align}
and
\begin{align}
\Omega_D=&\{\mathbf{y}:(\Sigma^{-1}\mathbf{e}_p)^\top\Sigma^{-1/2}_{inv}Q\mathbf{y}>P_0\} \nonumber\\
=&\{\mathbf{y}:\lambda y_1>P_0\}.
\end{align}

In fact, $\mathbf{Q}$ in \eqref{rotation_matrix} always exists. It would be the multiplication of $p-1$ rotation matrices which are extended from the 2-dimensional rotation matrix. For instance, $\mathbf{Q}=\mathbf{Q}_1\mathbf{Q}_2\cdots \mathbf{Q}_{p-1}$ with
\begin{equation}\label{matrix_example}
\mathbf{Q}_{p-1}=
\left[
\begin{array}{ccccc}
 1& & & &\\
 & 1& & &\\
& &\ddots & &\\
 &  & & \cos\tau_{p-1} &\sin\tau_{p-1}\\
 & & & -\sin\tau_{p-1}&\cos\tau_{p-1}
\end{array}
\right],
\end{equation}
and $\mathbf{Q}_{p-1}[w_1,\cdots,w_{p-1},w_p]^\top=[w_1,\cdots,\lambda_{p-1},0]^\top$. All of the remaining elements in \eqref{matrix_example} are zero and there is no need to calculate $\tau_{j},j=1,\cdots,p-1$. It is trivial to obtain that $\lambda_{p-1}=\sqrt{w_{p-1}^2+w_{p}^2}$. With the similar procedure, we can obtain $\lambda=\Vert(\Sigma^{-1}\mathbf{e}_p)^\top\Sigma^{-1/2}_{inv}\Vert$. Note that $\mathbf{Q}_j,j=1,2,\cdots,p-1$ and $\mathbf{Q}$ are orthogonal matrices. Therefore, we have $\mathbf{x}^\top\mathbf{x}=\mathbf{y}^\top(\mathbf{Q}^\top \mathbf{Q})\mathbf{y}=\mathbf{y}^\top\mathbf{y}$ and $\det(\mathbf{Q})=1$. Then, \eqref{P_e_Omega_D} becomes
\begin{align}
\hat{P}_e=\int_{\Omega_S}\int_{P_0\lambda^{-1}}^{+\infty}\frac{(1-\rho)k_2\det\left(\Sigma^{-1/2}_{inv}\right)}{\left(1+(y_1^2+\Vert\mathbf{y}_1\Vert^2)/\alpha\right)^\frac{\alpha+p}{2}}dy_1d\mathbf{y}_1,
\end{align}
where $\mathbf{y}_1=[y_2,y_3,\cdots,y_p]^\top$, $\Omega_S$ is the $(p-1)$-dimensional infinite space. We further apply spherical coordinate transform to simplify the integration as follows,
\begin{align}\label{P_e_polar_coordinate}
\hat{P}_e=&\int_{0}^{+\infty}\int_{P_0\lambda^{-1}}^{+\infty}\frac{(1-\rho)k_2\det\left(\Sigma^{-1/2}_{inv}\right)r^{p-2}}{\left(1+(y_1^2+r^2)/\alpha\right)^\frac{\alpha+p}{2}}dy_1dr \nonumber\\
&\times\underbrace{\int_{0}^{2\pi}\int_{0}^{\pi}\cdots\int_{0}^{\pi}\prod_{j=1}^{p-2}\left(\sin\beta_j\right)^{p-2-j}d\beta_1\cdots d\beta_{p-2}}_{\triangleq\Upsilon(p)}.
\end{align}

In the first part of the two-part paper, by calculating the marginal PDF of the GS model, we obtain that $\Upsilon(p)=\frac{(p-1)\pi^{p/2}}{2\Gamma(3/2)\Gamma((p+1)/2)}$ and therefore,
\begin{align}
\hat{P}_e=&\frac{(p-1)\pi^{\frac{p}{2}}(1-\rho)k_2\det\left(\Sigma^{-1/2}_{inv}\right)}{2\Gamma(\frac{3}{2})\Gamma(\frac{p+1}{2})}\nonumber\\
&\times\int_{P_0\lambda^{-1}}^{+\infty}\int_{0}^{+\infty}\frac{r^{p-2}}{\left(1+(y_1^2+r^2)/\alpha\right)^\frac{\alpha+p}{2}}drdy_1.
\end{align}

Based on \eqref{integral_identity} and \eqref{linear_trans}, the inner integral of \eqref{P_e_polar_coordinate} can be obtained to be
\begin{align}
&\int_{0}^{+\infty}\frac{r^{p-2}}{\left(1+(y_1^2+r^2)/\alpha\right)^\frac{\alpha+p}{2}}dr\nonumber\\
=&\frac{\alpha^{\frac{\alpha+p}{2}}\Gamma(\frac{p+1}{2})\Gamma(\frac{\alpha+1}{2})}{(p-1)\Gamma(\frac{\alpha+p}{2})}(\alpha+y_1^2)^{-\frac{\alpha+1}{2}}.
\end{align}

With similar manipulation, we have
\begin{align}
&\int_{P_0\lambda^{-1}}^{+\infty}\frac{\alpha^{\frac{\alpha+p}{2}}\Gamma(\frac{p+1}{2})\Gamma(\frac{\alpha+1}{2})}{(p-1)\Gamma(\frac{\alpha+p}{2})}(\alpha+y_1^2)^{-\frac{\alpha+1}{2}}dy_1\nonumber\\
=&\frac{\alpha^{\frac{p-1}{2}}\Gamma(\frac{p+1}{2})\Gamma(\frac{\alpha+1}{2})}{(p-1)\Gamma(\frac{\alpha+p}{2})}\bigg[\frac{\sqrt{\alpha}\Gamma(\frac{3}{2})\Gamma(\frac{\alpha}{2})}{\Gamma(\frac{\alpha+1}{2})}-\frac{ P_0}{\lambda}\nonumber\\
&\times{_{2}F_1}\bigg(\frac{1}{2},\frac{\alpha+1}{2};\frac{3}{2};-\frac{P_0^2}{\alpha\lambda^{2}}\bigg)\bigg].
\end{align}

Consequently, the theoretical expression of $\hat{P}_e$ is
\begin{align}\label{P_e_expression}
\hat{P}_e=&\frac{1-\rho}{2}\bigg[1-\frac{P_0\Gamma(\frac{\alpha+1}{2})}{\lambda\sqrt{\alpha}\Gamma(\frac{3}{2})\Gamma(\frac{\alpha}{2})}{_{2}F_1}\bigg(\frac{1}{2},\frac{\alpha+1}{2};\frac{3}{2};-\frac{P_0^2}{\alpha\lambda^{2}}\bigg)\bigg].
\end{align}

\section{Proof of theorem \ref{theorem_5}}\label{appendix_theorem_5}
\setcounter{equation}{0}
\renewcommand{\theequation}{C.\arabic{equation}}
As mentioned in section \ref{subsection_performance_bound_MSK}, a simple upper BER bound can be given as follows,
\begin{align}
P_{e,MSK}\leq P\big(f_M(\mathbf{n}_1+\Delta\mathbf{s}_{\theta})>f_M(\mathbf{n}_1)\big)\triangleq P^U_{e,MSK}.
\end{align}

By replacing $2A\mathbf{e}_p$ by $\Delta\mathbf{s}_{\theta}$, the integral region is the same as $\Omega_B$. By omitting the redundant derivation,  the upper bound given $\Delta\mathbf{s}_{\theta}$ is
\begin{align}\label{upper_bound_MSK_given_s_i}
P^U_{e,MSK}(\Delta\mathbf{s}_{\theta})=&\frac{1-\rho}{2}\bigg[1-\frac{\delta_{\theta}^B\Gamma(\frac{\alpha+1}{2})}{\sqrt{\alpha}\Gamma(\frac{3}{2})\Gamma(\frac{\alpha}{2})}\nonumber\\
&\times{_{2}F_1}\bigg(\frac{1}{2},\frac{\alpha+1}{2};\frac{3}{2};-\frac{(\delta_{\theta}^B)^2}{\alpha}\bigg)\bigg],
\end{align}
where
\begin{align}
\delta_{\theta}^B=\left|\frac{-\Delta\mathbf{s}_{\theta}^{\top}\Sigma^{-1}\Delta\mathbf{s}_{\theta}}{2\Vert\Delta\mathbf{s}_{\theta}^{\top}\Sigma^{-1}\Sigma^{-1/2}_{inv}\Vert}\right|.
\end{align}

Note that there are 4 kinds of $\Delta\mathbf{s}_{\theta}$ for MSK modulation and without loss of generality, we assume they are uniform distributed so that $P^U_{e,MSK}=\frac{1}{4}\sum_{\theta\in\Omega_{\theta}}P_{e,MSK}^U(\Delta\mathbf{s}_{\theta})$.

For the lower bound, we mainly focus on the first part of $P_{J,p}^L$ and the second part can be calculated similarly. Substituting $2A\mathbf{e}_p$ by $\Delta\mathbf{s}_{\theta}$ in $\Omega_A$, we can obtain $\delta_{\theta}^A=\Vert\Delta\mathbf{s}_{\theta}\Vert/2$. Let $\delta_{\theta}=\max\{\delta_{\theta}^A,\delta_{\theta}^B\}$. To solve $P_{J,p}^L$, we first consider that
\begin{align}
&\int_{\delta_{\theta}}^{2\delta_{\theta}}T(n_1)\int_{2\delta_{\theta}-n_1}^{+\infty}T(n_2)dn_2dn_1\nonumber\\
=&q_1^2\underbrace{\int_{\delta_{\theta}}^{2\delta_{\theta}}\frac{1}{(\alpha+n_1^2)^{\frac{\alpha+1}{2}}}\int_{2\delta_{\theta}-y_1}^{+\infty}\frac{dn_2dn_1}{(\alpha+n_2^2)^{\frac{\alpha+1}{2}}}}_{\triangleq P_{\delta}^A(\Delta\mathbf{s}_{\theta})},
\end{align}
where
\begin{align}
q_1=\frac{\alpha^{\frac{\alpha+p}{2}}(1-\rho)k_2\Upsilon(p)\sqrt{\det\left(\Sigma\right)}\Gamma(\frac{p+1}{2})\Gamma(\frac{\alpha+1}{2})}{(p-1)\Gamma(\frac{\alpha+p}{2})}.
\end{align}

The detailed derivation can be found in the appendix \ref{appendix_theorem_3}. For simplicity, we denote
\begin{align}\label{G_x}
G(x)=&\int (\alpha+x^2)^{-\frac{\alpha+1}{2}}dx\nonumber\\
=&x\alpha^{-\frac{\alpha+1}{2}}{_{2}F_1}\bigg(\frac{1}{2},\frac{\alpha+1}{2};\frac{3}{2};-\frac{x^2}{\alpha}\bigg),
\end{align}
\begin{align}\label{Gtilde_x}
\tilde{G}(x)=&\int G(x)dx=\alpha^{-\frac{\alpha+1}{2}}\Bigg[\frac{\alpha-\alpha\left(1+x^2/\alpha\right)^{\frac{1-\alpha}{2}}}{1-\alpha}\nonumber\\
&+x^2{_{2}F_1}\bigg(\frac{1}{2},\frac{\alpha+1}{2};\frac{3}{2};-\frac{x^2}{\alpha}\bigg)\Bigg].
\end{align}

Consequently,
\begin{align}\label{P_delta_1}
&P_{\delta}^A(\Delta\mathbf{s}_{\theta})\nonumber\\
=&\int_{\delta_{\theta}}^{2\delta_{\theta}}\frac{G(+\infty)-G(2\delta_{\theta}-n_1)}{(\alpha+n_1^2)^{\frac{\alpha+1}{2}}}dn_1\nonumber\\
=&G(+\infty)[G(2\delta_{\theta})-G(\delta_{\theta})]-\int_{\delta_{\theta}}^{2\delta_{\theta}}\frac{G(2\delta_{\theta}-n_1)}{(\alpha+n_1^2)^{\frac{\alpha+1}{2}}}dn_1.
\end{align}

Even though the remaining integral in \eqref{P_delta_1} has no closed expression, its approximation can be obtained based on the property of $G(x)$. We have
\begin{align}
&\int_{\delta_{\theta}}^{2\delta_{\theta}}\frac{G(2\delta_{\theta}-n_1)}{(\alpha+n_1^2)^{\frac{\alpha+1}{2}}}dn_1\nonumber\\
=&G(n_1)G(2\delta_{\theta}-n_1)\big|^{n_1=2\delta_{\theta}}_{n_1=\delta_{\theta}}-\int_{\delta_{\theta}}^{2\delta_{\theta}}\frac{G(n_1)dn_1}{(\alpha+(2\delta_{\theta}-n_1)^2)^{\frac{\alpha+1}{2}}}\nonumber\\
=&[G(\delta_{\theta})]^2-\frac{\tilde{G}(2\delta_{\theta})}{\alpha^{\frac{\alpha+1}{2}}}+\frac{\tilde{G}(\delta_{\theta})}{(\alpha+\delta_{\theta}^2)^{\frac{\alpha+1}{2}}}+(\alpha+1)\int_{\delta_{\theta}}^{2\delta_{\theta}}\tilde{G}(n_1)\nonumber\\
&\times(2\delta_{\theta}-n_1)(\alpha+(2\delta_{\theta}-n_1)^2)^{-\frac{\alpha+3}{2}}dn_1.
\end{align}

As $\tilde{G}(x)$ is too complicated to calculate, we further approximate it by some simple function. Actually, $\tilde{G}(x)$ is a convex function on $x\in[0,+\infty)$ since $\frac{\partial G(x)}{\partial x}$ is always larger than 0 on $[0,+\infty)$. In addition, $G(x)$ monotonically increases and has asymptotic behavior when $x\rightarrow+\infty$, i.e., $\lim\limits_{x\rightarrow+\infty}G(x)=\alpha^{-\frac{\alpha}{2}}\frac{\Gamma(3/2)\Gamma(\alpha/2)}{\Gamma((\alpha+1)/2)}$. Therefore, $\tilde{G}(x)$ can be approximated and upper bounded by an affine function. For the finite range $[a,b],a\geq0$, we have
\begin{align}\label{affine_1}
\tilde{G}(x)\leq\frac{\tilde{G}(b)-\tilde{G}(a)}{b-a}x+\frac{\tilde{G}(a)b-\tilde{G}(b)a}{b-a},x\in[a,b].
\end{align}

As for the infinite range $[a,+\infty),a>0$, the slope of $\tilde{G}(x)$ is always less than $G(+\infty)$ and consequently,
\begin{align}\label{affine_2}
\tilde{G}(x)\leq\frac{\Gamma(\frac{3}{2})\Gamma(\frac{\alpha}{2})}{\Gamma(\frac{\alpha+1}{2})}(x-a)+\tilde{G}(a),x\in[a,+\infty).
\end{align}

Based on the \eqref{affine_1}, we can obtain that
\begin{align}
&\int_{\delta_{\theta}}^{2\delta_{\theta}}\tilde{G}(n_1)(2\delta_{\theta}-n_1)(\alpha+(2\delta_{\theta}-n_1)^2)^{-\frac{\alpha+3}{2}}dn_1\nonumber\\
\leq&\int_{\delta_{\theta}}^{2\delta_{\theta}}\left[\frac{\tilde{G}(2\delta_{\theta})-\tilde{G}(\delta_{\theta})}{\delta_{\theta}}n_1+2\tilde{G}(\delta_{\theta})-\tilde{G}(2\delta_{\theta})\right]\nonumber\\
&\times(2\delta_{\theta}-n_1)(\alpha+(2\delta_{\theta}-n_1)^2)^{-\frac{\alpha+3}{2}}dn_1.
\end{align}

Based on the \eqref{integral_identity} and with some manipulations, we have
\begin{align}
&\int(an_1+b)(2\delta_{\theta}-n_1)(\alpha+(2\delta_{\theta}-n_1)^2)^{-\frac{\alpha+3}{2}}dn_1\nonumber\\
=&(2a\delta_{\theta}+b)\left[\alpha^{-\frac{\alpha+1}{2}}-\left(\alpha+(n_1-2\delta_{\theta})^2\right)^{-\frac{\alpha+1}{2}}\right]+\alpha^{-\frac{\alpha+1}{2}}\nonumber\\
&\times\frac{a(\alpha+1)}{3\alpha}(n_1-2\delta_{\theta})^3{_{2}F_1}\bigg(\frac{3}{2},\frac{\alpha+3}{2};\frac{5}{2};-\frac{(n_1-2\delta_{\theta})^2}{\alpha}\bigg).
\end{align}

By combining the previous analysis, the lower bound of the $P_{\delta}$ given $\Delta\mathbf{s}_{\theta}$ can be given as
\begin{align}
&P_{\delta}^A(\Delta\mathbf{s}_{\theta})\nonumber\\
=&G(+\infty)[G(2\delta_{\theta})-G(\delta_{\theta})]-[G(\delta_{\theta})]^2-\frac{\tilde{G}(\delta_{\theta})}{(\alpha+\delta_{\theta}^2)^{\frac{\alpha+1}{2}}}\nonumber\\
&+\frac{\tilde{G}(2\delta_{\theta})}{\alpha^{\frac{\alpha+1}{2}}}-\alpha^{-\frac{\alpha+1}{2}}(2a_1\delta_{\theta}+b_1)\left[1-\left(1+\frac{\delta_{\theta}^2}{\alpha}\right)^{-\frac{\alpha+1}{2}}\right]\nonumber\\
&+\frac{a_1\delta_{\theta}^3(\alpha+1)}{3\alpha^{\frac{\alpha+3}{2}}}{_{2}F_1}\bigg(\frac{3}{2},\frac{\alpha+3}{2};\frac{5}{2};-\frac{\delta_{\theta}^2}{\alpha}\bigg),
\end{align}
where $a_1=[\tilde{G}(2\delta_{\theta})-\tilde{G}(\delta_{\theta})]/\delta_{\theta}$ and $b_1=2\tilde{G}(\delta_{\theta})-\tilde{G}(2\delta_{\theta})$. Subsequently, the integral
\begin{align}
\int_{\delta_{\theta}}^{2\delta_{\theta}}T(n_1)\int_{2\delta_{\theta}-n_1}^{+\infty}\frac{\rho}{2\gamma_g\sqrt{\pi}}\exp\left(-\frac{n_2^2}{4\gamma_g^2}\right)dn_2dn_1,\nonumber
\end{align}
can also be calculated to get
\begin{align}
&\int_{\delta_{\theta}}^{2\delta_{\theta}}T(n_1)\int_{2\delta_{\theta}-n_1}^{+\infty}\frac{\rho}{2\gamma_g\sqrt{\pi}}\exp\left(-\frac{n_2^2}{4\gamma_g^2}\right)dn_2dn_1\nonumber\\
=&\frac{\alpha^{\frac{\alpha+p}{2}}\Gamma(\frac{p+1}{2})\Gamma(\frac{\alpha+1}{2})}{(p-1)\Gamma(\frac{\alpha+p}{2})}\Bigg\{\sqrt{\pi}\gamma_g[G(+\infty)-G(\delta_{\theta})]+\sqrt{\pi}\gamma_g\nonumber\\
&\times G(\delta_{\theta})\left[1-Q\left(\frac{\sqrt{2}\delta_{\theta}}{2\gamma_g}\right)\right]-\tilde{G}(2\delta_{\theta})+\tilde{G}(\delta_{\theta})\nonumber\\
&\times\exp\left(-\frac{\delta_{\theta}^2}{4\gamma_g^2}\right)+(2a_1\delta_{\theta}+b_1)-\exp\left(-\frac{\delta_{\theta}^2}{4\gamma_g^2}\right)(a_1\delta_{\theta}\nonumber\\
&+b_1)-a_1\sqrt{\pi}\gamma_g \left[1-Q\left(\frac{\sqrt{2}\delta_{\theta}}{2\gamma_g}\right)\right]\Bigg\}\triangleq P_{\delta}^B(\Delta\mathbf{s}_{\theta}).
\end{align}

Based on \eqref{affine_2}, the second part of the integration in \eqref{joint_expression_high_dimension} can be similarly derived. Then, the theoretical lower bound given  $\Delta\mathbf{s}_{\theta}$ can be expressed as
\begin{align}\label{lower_bound_given_delta_s_i}
&P^L_{e,MSK}(\Delta\mathbf{s}_{\theta})\nonumber\\
=&q_1^2\big\{P_{\delta}^A(\Delta\mathbf{s}_{\theta})+G(+\infty)[G(+\infty)-G(2\delta_{\theta})]\big\}+q_2\nonumber\\
&\times\big\{P_{\delta}^B(\Delta\mathbf{s}_{\theta})+G(+\infty)[G(+\infty)-G(2\delta_{\theta})]\big\},
\end{align}
where
\begin{align}
q_2=\frac{q_1\rho}{2\gamma_g\sqrt{\pi}}.
\end{align}

Therefore, Theorem \ref{theorem_5} could be concluded.
\end{appendices}

\footnotesize
\bibliographystyle{IEEEtran}
\bibliography{ref}

\begin{thebibliography}{10}
\providecommand{\url}[1]{#1}
\csname url@samestyle\endcsname
\providecommand{\newblock}{\relax}
\providecommand{\bibinfo}[2]{#2}
\providecommand{\BIBentrySTDinterwordspacing}{\spaceskip=0pt\relax}
\providecommand{\BIBentryALTinterwordstretchfactor}{4}
\providecommand{\BIBentryALTinterwordspacing}{\spaceskip=\fontdimen2\font plus
\BIBentryALTinterwordstretchfactor\fontdimen3\font minus
  \fontdimen4\font\relax}
\providecommand{\BIBforeignlanguage}[2]{{%
\expandafter\ifx\csname l@#1\endcsname\relax
\typeout{** WARNING: IEEEtran.bst: No hyphenation pattern has been}%
\typeout{** loaded for the language `#1'. Using the pattern for}%
\typeout{** the default language instead.}%
\else
\language=\csname l@#1\endcsname
\fi
#2}}
\providecommand{\BIBdecl}{\relax}
\BIBdecl

\bibitem{paper2}
P.~Chen, Y.~Rong, S.~Nordholm, Z.~He, and A.~J. Duncan, ``Joint channel
  estimation and impulsive noise mitigation in underwater acoustic ofdm
  communication systems,'' \emph{IEEE Trans. Wireless Commun.}, vol.~16, no.~9,
  pp. 6165--6178, Sept. 2017.

\bibitem{paper1}
U.~Epple and M.~Schnell, ``Advanced blanking nonlinearity for mitigating
  impulsive interference in ofdm systems,'' \emph{IEEE Trans. Veh. Technol.},
  vol.~66, no.~1, pp. 146--158, Jan. 2017.

\bibitem{paper20}
T.~Qi, W.~Huang, J.~Wang, and Q.~Peng, ``Viterbi demodulation of msk signal
  under both impulsive noise and gaussian white noise,'' in \emph{2023 IEEE
  98th Vehicular Technology Conference (VTC2023-Fall)}, 2023, pp. 1--5.

\bibitem{paper22}
H.~Qu, X.~Xu, J.~Zhao, F.~Yan, and W.~Wang, ``A robust hyperbolic tangent-based
  energy detector with gaussian and non-gaussian noise environments in
  cognitive radio system,'' \emph{IEEE Syst. J.}, vol.~14, no.~3, pp.
  3161--3172, Sept. 2020.

\bibitem{paper23}
C.~Chen, W.~Xu, Y.~Pan, H.~Zhu, and J.~Wang, ``Rank correlation based detection
  of known signals in middleton's class-a noise,'' \emph{IEEE Signal Process
  Lett}, vol.~28, pp. 1988--1992, Sept. 2021.

\bibitem{paper7}
D.~Middleton, ``Non-gaussian noise models in signal processing for
  telecommunications: new methods an results for class a and class b noise
  models,'' \emph{IEEE Trans. Inf. Theory}, vol.~45, no.~4, pp. 1129--1149,
  May. 1999.

\bibitem{paper8}
M.~Shao and C.~Nikias, ``Signal processing with fractional lower order moments:
  stable processes and their applications,'' \emph{Proc. IEEE}, vol.~81, no.~7,
  pp. 986--1010, July. 1993.

\bibitem{paper18}
J.~Gonzalez and G.~Arce, ``Weighted myriad filters: a robust filtering
  framework derived from alpha-stable distributions,'' in \emph{1996 IEEE
  International Conference on Acoustics, Speech, and Signal Processing
  Conference Proceedings}, vol.~5, 1996.

\bibitem{paper24}
S.~Kalluri and G.~Arce, ``Robust frequency-selective filtering using weighted
  myriad filters admitting real-valued weights,'' \emph{IEEE Trans. Signal
  Process}, vol.~49, no.~11, pp. 2721--2733, Nov. 2001.

\bibitem{paper25}
J.~M. Ramirez and J.~L. Paredes, ``Recursive weighted myriad based filters and
  their optimizations,'' \emph{IEEE Trans. Signal Process}, vol.~64, no.~15,
  pp. 4027--4039, Aug. 2016.

\bibitem{paper4}
D.~A. Chrissan and A.~C. Fraser-Smith, ``A clustering poisson model for
  characterizing the interarrival times of sferics,'' \emph{Radio Sci},
  vol.~38, Aug. 2003.

\bibitem{paper9}
A.~Mahmood and M.~Chitre, ``Modeling colored impulsive noise by markov chains
  and alpha-stable processes,'' in \emph{OCEANS 2015 - Genova}, 2015, pp. 1--7.

\bibitem{paper17}
------, ``Robust communication in bursty impulsive noise and rayleigh block
  fading,'' in \emph{WUWNet '16: Proceedings of the 11th International
  Conference on Underwater Networks \& Systems}, Oct. 2016, pp. 1--7.

\bibitem{paper26}
D.~Fertonani and G.~Colavolpe, ``On reliable communications over channels
  impaired by bursty impulse noise,'' \emph{IEEE Trans. Commun.}, vol.~57,
  no.~7, pp. 2024--2030, July. 2009.

\bibitem{paper14}
G.~Sureka and K.~Kiasaleh, ``Sub-optimum receiver architecture for awgn channel
  with symmetric alpha-stable interference,'' \emph{IEEE Trans. Commun.},
  vol.~61, no.~5, pp. 1926--1935, May. 2013.

\bibitem{book3}
J.~Proakis and M.~Salehi, \emph{Digital Communications}.\hskip 1em plus 0.5em
  minus 0.4em\relax New York: McGraw Hill, 2015.

\bibitem{book4}
D.~Zwillinger, V.~Moll, I.~Gradshteyn, and I.~Ryzhik, Eds., \emph{Table of
  Integrals, Series, and Products (Eighth Edition)}.\hskip 1em plus 0.5em minus
  0.4em\relax Boston: Academic Press, 2014.

\bibitem{book7}
S.~J. Taylor, \emph{Introduction to Measure and Integration}.\hskip 1em plus
  0.5em minus 0.4em\relax Cambridge: Cambridge University Press, 1973.

\bibitem{book2}
G.~Samorodnitsky, M.~S.~T. Chapman, and Hall, \emph{Stable non-Gaussian random
  processes: Stochastic models with infinite variance}.\hskip 1em plus 0.5em
  minus 0.4em\relax New York: Chapman \& Hall, 1994.

\bibitem{paper21}
G.~Forney, ``The viterbi algorithm,'' \emph{Proc. IEEE}, vol.~61, no.~3, pp.
  268--278, 1973.

\end{thebibliography}

\end{document}